\documentclass[12pt]{article}

\usepackage[T1]{fontenc}
\usepackage[utf8]{inputenc}
\usepackage{lmodern}
\usepackage{microtype}
\usepackage{arxiv}

\usepackage{amsthm}
\usepackage{natbib}
\usepackage{multirow}
\usepackage{booktabs}
\usepackage{array}
\usepackage{float}
\newfloat{algorithm}{tbp}{loa}
\floatname{algorithm}{Algorithm}

\usepackage{mdframed}
\usepackage{tikz}
\usetikzlibrary{patterns,positioning,shapes,arrows.meta}
\usepackage{xcolor}

\usepackage{setspace}
\usepackage{caption}
\usepackage{geometry}

\usepackage{xr-hyper}
\usepackage{hyperref}
\IfFileExists{xurl.sty}{\usepackage{xurl}}{}

\AtBeginEnvironment{table}{\normalsize\doublespacing}
\AtBeginEnvironment{algorithm}{\normalsize\doublespacing}

\makeatletter
\def\maxwidth{\ifdim\Gin@nat@width>\linewidth\linewidth\else\Gin@nat@width\fi}
\def\maxheight{\ifdim\Gin@nat@height>\textheight\textheight\else\Gin@nat@height\fi}
\makeatother
\setkeys{Gin}{width=\maxwidth,height=\maxheight,keepaspectratio}

\hypersetup{
  pdftitle={Simultaneous False Discovery Proportion Inference under Sensitivity Models},
  pdfauthor={Mengqi Lin; Colin Fogarty},
  pdfkeywords={Causal Inference; Closed Testing; Exploratory Data Analysis; Randomization Inference; Sensitivity Analysis; Simultaneous Inference},
  hidelinks
}
\theoremstyle{plain}
\newtheorem{theorem}{Theorem}

\newtheorem{proposition}{Proposition}

\theoremstyle{definition}

\newtheorem{example}{Example}
\newtheorem{remark}{Remark}

\begin{document}

\title{\bfseries Simultaneous Inference for False Discovery Proportions under Sensitivity Models for Observational Studies}
\author{Mengqi Lin and Colin B. Fogarty%
\\
Department of Statistics, University of Michigan}
  
\date{}
\maketitle

\begin{abstract}
\singlespacing
We provide an approach to exploratory data analysis in observational studies with a single intervention and multiple endpoints. In such settings, the researcher would like to explore evidence for actual treatment effects while accounting for both false discoveries and the potential impact of unmeasured confounding. For any candidate subset of hypotheses about these outcomes, we provide sensitivity sets for the proportion of the hypotheses within the subset which are true. The resulting sensitivity statements are valid simultaneously over all possible choices for the rejected set, allowing the researcher to search for promising subsets of hypotheses that maintain a large estimated fraction of true discoveries even if hidden bias is present. The approach is well suited to sensitivity analysis, as conclusions that some fraction of outcomes are affected by the treatment exhibit larger robustness to unmeasured confounding than findings that any particular outcome is affected. For Rosenbaum's sensitivity model in matched observational studies, we show how a sequence of integer programs, in tandem with screening steps, facilitate the efficient computation of the required sensitivity sets. We illustrate the practical utility of our method through both simulation studies and a data example on the long-term impacts of childhood abuse.\end{abstract}


\section{Introduction}

In observational studies, lack of control over the assignment of interventions leaves causal conclusions susceptible to bias from unobserved covariates. A sensitivity analysis identifies the strength of unmeasured confounding that would be required to overturn the findings of the study. Observational studies can vary considerably in their robustness to hidden bias; see, for instance, Table 4.10 of \citet{Rosenbaum2002}. At one extreme, the study of smoking and lung cancer by \cite{ham64} produced evidence of a causal effect that may only be reversed by unmeasured confounders with such a large impact on the chances of smoking as to be viewed implausible in practice. More common however are studies where any particular finding can be overturned by a strength of hidden bias that is harder to dismiss out of hand. In short, small treatment effects cannot generally be distinguished from small amounts of hidden bias. This inspires both the development of informative inferential targets which may exhibit improved robustness to hidden bias; and the provision of techniques allowing practitioners to actively search for robust findings. 

The false discovery proportion (FDP) for a set of hypotheses $\mathcal{R}$ is the fraction of hypotheses within $\mathcal{R}$ that are actually true \citep{benjamini1995controlling}. Methods controlling the false discovery rate (FDR) provide procedures for choosing the set $\mathcal{R}$ such that the FDP is controlled at a pre-specified rate in expectation. \citet{goemansolari} provide an alternative notion of control tied to the FDP more closely aligned with exploratory research. Rather than prescribing a particular approach for choosing the set of hypotheses $\mathcal{R}$, they instead describe how one can provide simultaneously valid confidence sets for the FDP of \emph{any} chosen subset of hypotheses $\mathcal{R}$ based upon the output of a closed testing procedure \citep{marcus1976closed}. Their approach, and the simultaneous error control it provides, allows researchers to actively search for promising sets of hypotheses where the upper confidence limit for the FDP is small (equivalently, where the lower confidence bound for the number of false hypotheses in the set is large). 

In this work, we develop simultaneously valid sensitivity sets for the false discovery proportion in the presence of multiple hypotheses on treatment effects while allowing for unmeasured confounding of varying strengths. For any given set of hypotheses $\mathcal{R}$, the changepoint levels produced in the sensitivity analysis quantify how strong the hidden bias would need to be before the upper confidence bound on the fraction of null effects exceeds a given threshold. This allows researchers to assess the robustness of the finding that certain fraction of hypotheses are false among \emph{any} chosen subset $\mathcal{R}$. As will be discussed, findings that a given fraction of hypotheses are false are more robust to hidden bias than the finding that any particular hypothesis is false. This can be useful for several reasons. For instance, if one was looking at the impact of childhood abuse on risk-seeking behavior as an adult, knowing that the finding that half of the outcomes reflective of risk-seeking behavior were affected is robust to hidden bias would still be valuable even if one could not identify which particular outcomes were affected. 
Furthermore, simultaneous validity facilitates exploratory
comparisons of subsets of outcomes on the basis of their collective resilience to
unmeasured confounding. Because the confidence statements hold simultaneously
for every subset, researchers may examine multiple scientifically
meaningful subsets after observing the data and still provide a valid
sensitivity statement to the subset ultimately chosen. This helps researchers identify subsets of outcomes for which the claim
that at least a specified fraction of the outcomes are affected remains robust
to stronger hidden bias. Such findings may then be prioritized for replication
or further investigation.

At the heart of the statistical appeal of simultaneous FDP control within a sensitivity analysis is the natural occurrence of \emph{non-consonant rejections} — where intersections of hypotheses may be rejected without rejecting any individual hypothesis. \citet{goemansolari} illustrate how non-consonant rejections, often viewed as wasteful when using closed testing methods for familywise error control, provide additional information in the context of FDP control. As we will show, such rejections arise frequently in sensitivity analysis, a fact that has been leveraged to improve the power for testing individual hypotheses in the existing literature on sensitivity analysis with multiple endpoints \citep{ros16, fogarty2016sensitivity, zheng2024}. In the presence of non-consonant rejections, although we cannot pinpoint exactly which hypotheses are false, focusing on the FDP allows us to determine how many of the hypotheses are false at a given confidence level. This enables us to extract valuable information that would otherwise be lost if we focused solely on individual hypotheses.

\subsection{Motivating example: long-term impacts of childhood abuse}\label{sec:example} 
Childhood physical and emotional abuse (CPEA) has been linked in the literature to a broad range of adverse adult outcomes, including psychosocial well-being \citep{herrenkohl2012prospective} and personality \citep{lee2017childhood},  mental health \citep{edwards2003relationship}, 
health behaviors such as drinking \citep{caetano2003association, anne2011child} and smoking \citep{kristman2013child}, physical health such as obesity \citep{danese2014childhood}, interpersonal relationships \citep{maneta2012links,paradis2019child}, and socio-economic attainment \citep{bunting2018association}. 
Datasets such as the Wisconsin Longitudinal Study (WLS), which collect information across many of these domains, offer an opportunity to explore which combinations of adverse outcomes are most plausibly causally affected by CPEA.

Exploring and discovering causal effects among these outcome variables introduces two main challenges. First, even if the data were the result of a randomized experiment, testing multiple outcomes simultaneously risks inflating the false discoveries if not properly controlled.  Second, while many methods exist to adjust for observed covariates, seemingly relevant confounders remain unobserved in this study.  For instance, a parent’s unreported mental health condition or history of arrests may influence both the likelihood of childhood abuse and outcomes in adulthood for their children, 
raising the possibility of unmeasured confounding.  Our proposal provides simultaneous confidence sets for the proportion of outcomes that are truly affected by childhood abuse for any subset of outcomes, and quantifies how robust these confidence statements are to varying strengths of hidden bias. Researchers can thus explore which collections of outcomes maintain a small FDP while allowing for hidden bias. As we demonstrate in Section~\ref{sec:application}, we remain confident that within certain combinations of outcomes—such as alcohol, closeness to spouse, anger and agreeableness, at least half of them are affected by the treatment at larger degrees of hidden bias than is required to overturn the findings on any of the particular outcomes. This evidence can help researchers prioritize these outcomes for further investigation, with greater confidence that they reflect true impacts of CPEA even in the presence of moderate unmeasured confounding.

The remainder of this paper is organized as follows. In Section~\ref{sec:causal_framework}, we introduce the general sensitivity framework with multiple outcomes, including the formal definition of the FDP.  We then present our main methodological contributions, showing how to construct simultaneously valid sensitivity sets for the FDP in Section~\ref{sec:SA_FDP}. Computational strategies and improvements are discussed in Section~\ref{sec:computation}. Simulation studies reflective of the many uses of our proposal are presented in Section~\ref{sec:simulation}, while Section~\ref{sec:application} applies our techniques to the CPEA example.

\section{Sensitivity models with multiple outcomes}\label{sec:causal_framework}

\subsection{Potential outcomes and multiple hypotheses}\label{subsec:causal_notation}
Consider an observational study with \(N\) units and \(K\) outcomes of interest. For unit \(i\in\{1,\ldots,N\}\), let \(Z_i\in\{0,1\}\) denote the treatment assignment indicator. For outcome \(k\in[K]:=\{1,\ldots,K\}\), let \(r_{Tik}\) and \(r_{Cik}\) denote the potential outcomes under treatment and control, respectively. Write \(\br_{Ti}=(r_{Ti1},\ldots,r_{TiK})^\top \in \R^K\), \(\br_{Ci}=(r_{Ci1},\ldots,r_{CiK})^\top \in \R^K\), and \(\bR_i=(R_{i1},\ldots,R_{iK})^\top\), where \(\bR_i=Z_i\br_{Ti}+(1-Z_i)\br_{Ci}\) \citep*{ney23,rub74} is the observed outcome vector. Let \(\bx_i\in\mathbb R^P\) denote observed pre-treatment covariates, and let \(u_i\) denote unobserved covariates that may affect treatment assignment and outcomes.
Let $\bZ=(Z_1,\ldots,Z_N)^\top$ be the \(N\)-dimensional binary vector of treatment assignments for all individuals. Similarly, let \(\mathbf R\), \(\br_T\), and \(\br_C\) be the \(N\times K\) matrices of observed outcomes and potential outcomes; analogously for \(\bx\) and \(\bu\).

For each outcome \(k\), let \(H_k\) denote a null hypothesis relating the treated and control potential outcomes for outcome \(k\). The formulation of \(H_k\) depends on the inferential framework and estimand under consideration. Common examples include the sharp null of no treatment effect for any individual for outcome $k$; and the weak null of no average effect for outcome $k$.
For the elementary hypotheses \(H_1,\ldots,H_K\),
for any subset \(\cI\subseteq[K]\), define the intersection null hypothesis $H_{\cI}=\bigcap_{k\in\cI} H_k.$
Let $\cH_0=\{k\in[K]:H_k\text{ is true}\}$ denote the set of true elementary null hypotheses. 
For any subset of outcomes \(\cR\subseteq[K]\), define $V(\cR)=|\cR\cap\cH_0|.$
The ratio ${V(\cR)}/\{|\cR|\vee 1\}$
is the fraction of true null hypotheses in \(\cR\). 
When \(\cR\) is interpreted as a rejection set, this quantity is the false discovery proportion. 
We maintain the terminology FDP throughout, although the inferential statements developed below apply simultaneously to all subsets \(\cR\subseteq[K]\), not only to rejection sets generated by a pre-specified multiple testing procedure.

\subsection{Sensitivity models as collections of null distributions}
\label{subsec:sensitivity_model}

In an observational study where treatment assignment and outcomes may depend on latent confounders, even after adjusting for the observed covariates \(\bx_i\), the null distributions of the test statistics are still not identified from the observed data. 
A sensitivity model at sensitivity parameter \(\Gamma\) posits a set of possible
hidden bias configurations compatible with the observed data, denoted by
\(\cP_\Gamma\), where \(\Gamma\) measures the strength of hidden bias. We write
\(\brho\in\cP_\Gamma\) for a generic hidden bias configuration. Each \(\brho\) induces a
probability measure \(\P_{\brho}\).

Suppose that for any given \(\brho\), a \(p\)-value \(p_{\cI,\brho}\),
valid either in finite samples or asymptotically, may be constructed for testing $H_\cI$; see the examples below and Section~\ref{sec:computation} for a concrete construction.
As \(\brho\) is unknown in practice due to its dependence on the unknown latent confounder $\bu$, inference at sensitivity level \(\Gamma\) is instead carried out under the least favorable hidden bias configuration in \(\cP_\Gamma\), namely the configuration that most challenges the inferential claim under consideration. For example, when testing only the elementary null $H_k$ 
one computes the
worst-case \(p\)-value 

\begin{equation}\label{eq:worst_pval}
    p_{k,\Gamma}^*
    =
    \max_{\brho\in\mathcal P_\Gamma}
    p_{k,\brho}.
\end{equation}

We take \(\Gamma=1\) to correspond to the NUC (no unmeasured confounding) setting, and larger values of
\(\Gamma\) to allow larger departures from NUC. That is, for
\(\Gamma_1\le\Gamma_2\), we have
\(\cP_{\Gamma_1}\subseteq\cP_{\Gamma_2}\).
Thus, increasing \(\Gamma\) enlarges the class of hidden bias configurations considered plausible, the least favorable inference becomes more conservative. 
As $\Gamma$ increases, there exists a \(\Gamma^*\), called \emph{sensitivity value} \citep{zhao19}, at which the statistical conclusion changes. This sensitivity value \(\Gamma^*\) serves as a measure of the robustness of the findings to hidden bias: a larger \(\Gamma^*\) indicates greater resilience to hidden bias, whereas a smaller \(\Gamma^*\) indicates that a modest strength of hidden bias would suffice to overturn the observed results. For instance, when testing only the elementary null $H_k$ at a given significance level \(\alpha\), the sensitivity value \(\Gamma_k^*\) corresponds to the smallest \(\Gamma\) at which the worst-case \(p\)-value exceeds \(\alpha\):

\[
    \Gamma_k^*
    =
    \inf
    \left\{
    \Gamma\geq1:
    p_{k,\Gamma}^*>\alpha
    \right\}.
\]
This general framework covers several sensitivity models in the literature; three representative examples are provided below.

\begin{example}[Rosenbaum's sensitivity model \citep{ros87}]\label{ex:rosenbaum}
Consider a matched observational study analyzed under the finite-population framework. 
Matching on observed covariates partitions the \(N\) units into matched sets
\(\mathcal M_1,\ldots,\mathcal M_B\). 
Let \(\cZ\) denote the event that the treatment assignment respects the matched design and
\(\cF=\{(\br_{Ti},\br_{Ci},\bx_i,u_i):i=1,\ldots,N\}\). In finite-population framework, where inference conditions upon \(\cF\), randomness arises solely from \(\bZ\) \citep{fis35}.
Let \(\pi_i=\P(Z_i=1\mid\cF)\), the model assumes that
\begin{equation}\label{eq:sensmodel}
\log\!\left(\frac{\pi_i}{1-\pi_i}\right)
=
\kappa(\bx_i)+u_i\log\Gamma,\qquad 0\le u_i\le 1.
\end{equation}
where $\kappa(\cdot)$ is an unknown function of the observed covariates. Therefore, for all \(i,i'\in\mathcal M_b\) with $\bx_i = \bx_i'$, the model bounds their treatment assignments by requiring \(1/\Gamma\le \{\pi_i(1-\pi_{i'})\}/\{\pi_{i'}(1-\pi_i)\}\le\Gamma\), due to their differences in $u$.
Equivalently, after conditioning on \(\cZ\), let
\(\varrho_i=\P(Z_i=1\mid\cF,\cZ)\) and
\(\brho=(\varrho_1,\ldots,\varrho_N)^\top\). Then \(\cP_\Gamma\) consists of
all \(\brho\) such that \(\varrho_i\ge0\),
\(\sum_{i\in\mathcal M_b}\varrho_i=1\), and
\(\Gamma^{-1}\le\varrho_i/\varrho_{i'}\le\Gamma\) for all
\(i,i'\in\mathcal M_b\).
When \(\Gamma=1\), all units in the same
matched set are equally likely to receive treatment, recovering NUC setting, whereas larger \(\Gamma\) allows greater within-set
selection bias.
\end{example}

\begin{example}[Marginal sensitivity model \citep{tan06,Dorn23}]
Consider an i.i.d. superpopulation setting with observed data
\((\bx_i,Z_i,\bR_i)\) and full data
\((\bx_i,u_i,Z_i,\br_{Ti},\br_{Ci})\). The model assumes latent unconfoundedness \((\br_{Ti},\br_{Ci})\indep Z_i\mid(\bx_i,u_i)\). Let
\(e(\bx_i)=\P(Z_i=1\mid\bx_i)\) be the nominal propensity score and
\(e_0(\bx_i,u_i)=\P(Z_i=1\mid\bx_i,u_i)\) be the true propensity score, the model assumes that
\[
    1/\sqrt{\Gamma}
    \le
    \frac{e_0(\bx_i,u_i)/(1-e_0(\bx_i,u_i))}
         {e(\bx_i)/(1-e(\bx_i))}
    \le
    \sqrt{\Gamma} .
\]
Equivalently, for each \(\bx_i\), \(e_0(\bx_i,u_i)\) must lie between
\(e(\bx_i)/[e(\bx_i)+\sqrt{\Gamma}\{1-e(\bx_i)\}]\) and
\(\sqrt{\Gamma} e(\bx_i)/[1-e(\bx_i)+\sqrt{\Gamma} e(\bx_i)]\).
Here the hidden bias configuration is \(\varrho_i\) = \(\text{logit}\{e(\bx_i)\}-\text{logit}\{e_0(\bx_i,u_i)\}\), whereas \(\cP_\Gamma\) consists
of all such configurations satisfying the above constraint and
compatible with the observed data. When
\(\Gamma=1\), \(e_0(\bx_i,u_i)=e(\bx_i)\), corresponding to NUC.
\end{example}

\begin{example}[Factor-structured partial-\(R^2\) sensitivity model
\citep{zheng2024}]
Consider an i.i.d. superpopulation setting with latent confounders \(u_i\). The model assumes latent unconfoundedness
\((\br_{Ti},\br_{Ci})\indep Z_i\mid(\bx_i,u_i)\) and factor-structured outcomes. Let
\(Z_i^\perp=Z_i-\E(Z_i\mid\bx_i)\) and $\bu_i^\perp=\bu_i-\E(\bu_i\mid\bx_i)$, then
\(\|\brho\|_2^2
=
{
\operatorname{cov}(Z_i^\perp,\bu_i^\perp)
\operatorname{cov}(\bu_i^\perp)^{-1}
\operatorname{cov}(\bu_i^\perp,Z_i^\perp)
}/{
\operatorname{var}(Z_i^\perp)
}\) is the linear partial \(R^2\) between treatment and the latent
factors after adjustment for \(\bx_i\).
Under the model and factor-confounding conditions of \citet{zheng2024}, the remaining unidentified bias can be parameterized by $\brho$, and $\cP_\Gamma
    =
    \left\{
        \brho:
        \|\brho\|_2^2\le 1-\Gamma^{-1}
    \right\}.$
Thus \(\Gamma=1\) forces \(\brho=0\), corresponding to no residual linear association between treatment and the latent factors, while
larger \(\Gamma\) permits a larger treatment--factor partial \(R^2\).
\end{example}

\section{Sensitivity sets for the false discovery proportion}\label{sec:SA_FDP}
\subsection{Confidence sets with known $\brho$}\label{subsec:FDP_method}
Suppose for the purpose of illustration that the hidden bias configuration
\(\brho\) is fixed.
To provide simultaneous confidence sets for the false discovery proportion, we consider the method of \citet{goemansolari}, which makes direct use of a \emph{closed testing procedure} \citep{marcus1976closed}. Closed testing requires the specification of a level-\( \alpha \) test for each intersection null hypothesis \( H_{\cI} \), called the \emph{local test} for $H_{\cI}$. The closed testing procedure rejects \( H_{\cI} \) if and only if all \( H_{\cJ} \) with which \( \cJ \supseteq \cI \) are rejected by their corresponding local tests. For a given $\brho$, we can obtain a local test for each $H_{\cI}$. We denote rejection by this local test, which depends on $\brho$, as \( \phi_{\cI,\brho} \), and further denote the rejection of \( H_{\cI} \) in the closed testing procedure under \(\brho\) by \( \tilde{\phi}_{\cI,\brho} \), then $\tilde{\phi}_{\cI,\brho} = \min_{\cJ \supseteq \cI} \phi_{\cJ,\brho}$.
Based on this, a \( 100(1-\alpha)\% \) confidence set for the FDP: $\set{
0, \;{1}/{|\cR|}, \;\ldots, \;{v_{\brho}(\cR)}/{|\cR|}}$
is valid simultaneously for any given set \( \cR \) \citep{goemansolari},
where the quantity \( v_{\brho}(\cR) \) (with its dependence on $\alpha$ suppressed) is defined as the size of the largest subset of \( \cR \) for which the corresponding intersection null hypothesis is \emph{not} rejected in the closed testing procedure:

\begin{equation}\label{eq:vR}
    v_{\brho}(\cR) = \max_{\cI \subseteq \cR} |\cI| \cdot (1 - \tilde{\phi}_{\cI, \brho}),
\end{equation}
where $1 - \tilde{\phi}_{\cI, \brho}$ indicates the failure of rejection of $H_{\cI}$ in the closed testing procedure.

The ease with which \( v_{\brho}(\cR) \) may be computed is determined in part by whether or not the closed testing procedure is \emph{consonant}. A closed testing procedure is consonant if any rejection of an intersection null hypothesis \( H_{\cI} \) leads to the rejection of at least one elementary hypothesis \( H_k \) within \( \cI \), i.e. any $H_{\cI}$ for which $\tilde{\phi}_{\cI,\brho} = 1$ implies the existence of at least one $k \in \cI$ such that $\tilde{\phi}_{k,\brho} = 1$.
For consonant procedures, \( v_{\brho}(\cR) \) can be easily obtained by counting the number of elementary hypotheses in \( \cR \) that are not rejected:
\begin{singlespace}
\begin{equation}\label{eq:vR_cons}
   v_{\brho}(\cR) = \sum_{k \in \cR} (1 - \tilde{\phi}_{k,\brho}).
\end{equation} 
\end{singlespace}
\noindent
In such cases, it suffices to determine the rejections of elementary hypotheses, which can be done efficiently for many closed testing procedures. In contrast, if the procedure is not consonant, computing \( v_{\brho}(\cR) \) generally requires access to the rejections of the entire closed testing tree, which includes all intersection null hypotheses; see Section~\ref{subsubsec:non_consonance} for additional discussion.

As an illustration, consider a Bonferroni-based closed testing procedure where the local tests reject \( H_{\cI} \) if the smallest \( p \)-value in \( \cI \) is less than \( \alpha / |\cI| \). Under a given \(\brho\), we may therefore first compute the \( p \)-values \( p_{k,\brho} \) for each elementary hypothesis \( H_k \) and obtain the local test:
\begin{equation}\label{eq:bonfrho}
    \phi_{\cI,\brho} = \mathbf{1}\left\{ \min_{k \in \cI} p_{k,\brho} \leq \frac{\alpha}{|\cI|} \right\}.
\end{equation}
The procedure based on these local tests is consonant. Therefore, obtaining $v_\brho(\cR)$ for any $\cR$ requires only the rejections of the elementary hypotheses, which can be efficiently determined using the Holm-Bonferroni procedure \citep{hol79}.



\subsection{The generalized sensitivity value}\label{sec:gSV}
As noted in Section~\ref{subsec:sensitivity_model}, knowing $\brho$ is unrealistic due to its dependence on the unobserved latent confounders.
We next describe how sensitivity analysis may provide simultaneous confidence sets for the false
discovery proportion at a posited strength of hidden bias $\Gamma$.
Since a larger value of $v_{\brho}(\mathcal{R})$ represents more potential false discoveries, the sensitivity analysis at a given \( \Gamma \), in this context, seeks to find the largest value of \( v_{\brho}(\mathcal{R}) \) over all allowable configurations \( \brho \in \mathcal{P}_\Gamma \). We denote this worst-case quantity as $v_{\Gamma}^*(\mathcal{R}) := \max_{\brho \in \mathcal{P}_\Gamma} v_{\brho}(\mathcal{R})$.
Based on this, we obtain the most conservative confidence set for FDP over all allowable $\adverrho$, which accounts for the most adverse unmeasured confounding effects permissible under the specified level \( \Gamma \). 
We refer to this set as \emph{sensitivity set}, and formalize it below.

\begin{theorem}[Simultaneous validity]
\label{thm:sensitivity_sets_valid}
Suppose that, for each \(\brho\), every
\(\phi_{\cI,\brho}\) is a level-\(\alpha\) test of \(H_{\cI}\) under
\(\P_{\brho}\). Let \(\brho_0\) denote the true hidden bias configuration
and suppose that
\(\brho_0\in\cP_{\Gamma_0}\). Then
\[
\P_{\brho_0}\left\{
    V(\cR)\leq v_\Gamma^*(\cR)
    \ \text{for every }\cR\subseteq[K]
\right\}
\geq 1-\alpha
\]
holds for every $\Gamma\geq\Gamma_0$.
Consequently, $\set{
        0,\frac{1}{|\cR|},\ldots,\frac{v_\Gamma^*(\cR)}{|\cR|}
    }$
defines a simultaneous \(100(1-\alpha)\%\) sensitivity set for the FDP
over all nonempty \(\cR\subseteq[K]\), for every $\Gamma\geq\Gamma_0$.
\end{theorem}

As \( \Gamma \) increases, the sensitivity model permits stronger hidden bias, so
\(v_{\Gamma}^*(\cR)\) is nondecreasing in \(\Gamma\). In other words, increasing $\Gamma$ can never result in a smaller upper bound on the number of false discoveries in the context of sensitivity analysis, and may increase this upper bound. Since
\(v_{\Gamma}^*(\cR)\) takes integer values in
\(\{0,1,\ldots,|\cR|\}\), for each \( r \in \{0, 1, \ldots, |\cR| - 1\} \), there exists a value of \( \Gamma \) at which \( v_{\Gamma}^*(\cR) \) increases from \( r \) to \( r + 1 \). Denote this changepoint value by \( \Gamma^*(\cR, r) \), then
\begin{equation}\label{eq:sensitivity_Gamma}
\Gamma^*(\cR,r)
=
\inf\left\{
    \Gamma\ge 1:
    v_{\Gamma}^*(\cR)>r
\right\}.
\end{equation}
We refer to \( \Gamma^*(\cR,r) \) as the \emph{generalized sensitivity value} for \(\cR\) at candidate number \(r\) of true null hypotheses. This definition
generalizes the sensitivity value of \citet{zhao19}, which was defined for a single
outcome, to arbitrary subsets of outcomes.
It provides a measure of the robustness of \( \cR \) against unmeasured confounding. Specifically, 
for any \(\Gamma<\Gamma^*(\cR,r)\), the sensitivity set implies \(V(\cR)\le r\), and hence at least \(|\cR|-r\) hypotheses in \(\cR\) are true discoveries. Therefore, a
larger value of \(\Gamma^*(\cR,r)\) indicates greater robustness of the claim that
at least \(|\cR|-r\) hypotheses in \(\cR\) are true discoveries.


To illustrate this concept, consider Figure~\ref{fig:sensitivity_intervals}. As \( \Gamma \) increases, $v_{\Gamma}^*(\cR)$ increases in a stepwise manner. 
Each value \(\Gamma^*(\cR,r)\) indicates the smallest amount of hidden bias at
which the upper confidence bound on \(V(\cR)\) can exceed \(r\). 
For example, \(\Gamma^*(\cR,0)=2\) implies that, for all \(\Gamma<2\) the $1-\alpha$ sensitivity set supports the
claim that \(V(\cR)=0\), i.e. that all hypotheses in \(\cR\) are true discoveries. Once
\(\Gamma\) reaches \(2\), the sensitivity set can include the possibility that
\(V(\cR)\ge1\), meaning that at least one hypothesis in \(\cR\) may be a true null.
In the matched observational study setting of Example~\ref{ex:rosenbaum}, this
corresponds to allowing the odds of treatment assignment for two individuals in the
same matched set to differ by a factor of \(2\).

\begin{figure}[h!]
    \centering
    \begin{tikzpicture}[xscale=2, yscale=2, scale = 0.8]
    \draw[->] (0.8,0) -- (4.8,0) node[right] {$\Gamma$};
    \draw[->] (0.8,-0.5) -- (0.8,2.5) node[above] {$v_{\Gamma}^*(\cR)$};
    
    \draw[ultra thick] (1,0) -- (2,0);
    \draw[ultra thick] (2,0) -- (2,1);
    \draw[ultra thick] (2,1) -- (3.7,1);
    \draw[ultra thick] (3.7,1) -- (3.7,2);
    \draw[ultra thick] (3.7,2) -- (4,2);
    
    \draw[dashed, gray, thick] (3.7,0) -- (3.7,2.2);
    
    \foreach \x/\label in {1/1,2/2,3/3,4/4}{
        \draw (\x,0) -- (\x,-0.05);
        \node[below=2pt, yshift=-1pt] at (\x,0) {\label};
    }

    \foreach \x in {2,3.7}{
        \draw[red, line width=0.9pt] 
            (\x-0.07,-0.07) -- (\x+0.07,0.07)
            (\x-0.07,0.07) -- (\x+0.07,-0.07);
    }

    \node[above=3pt, red, font=\small] at (2.5,-0.1) {$\Gamma^*(\cR,0)$};
    \node[above=3pt, red, font=\small] at (4.2,-0.1) {$\Gamma^*(\cR,1)$};
    
    \foreach \y/\label in {0/0,1/1,2/2}{
        \draw (0.8,\y) -- (0.7,\y) node[left] {\label};
    }
\end{tikzpicture}\caption{$v_{\Gamma}^*(\cR)$ as a function of $\Gamma$.}
    \label{fig:sensitivity_intervals}
\end{figure}

\begin{proposition}
\label{prop:collective_robustness}
For any nonempty subset \(\cR\subseteq[K]\),
\[
    \Gamma^*(\cR,|\cR|-1)
    \ge
    \max_{k\in\cR}\Gamma^*(\{k\},0).
\]
\end{proposition}

Proposition~\ref{prop:collective_robustness} implies that, the claim that some outcome in \(\cR\) is affected is more robust than the claim that any particular outcome in \(\cR\) is affected.

We now provide two practical applications of the generalized sensitivity value.

\paragraph{Application 1: quantifying robustness of a subset.}

For a scientifically chosen subset of outcomes \(\cR\), the generalized
sensitivity value \(\Gamma^*(\cR,r)\) quantifies the robustness of the claim
that at least \(|\cR|-r\) outcomes in \(\cR\) are affected. Indeed, for every
\(\Gamma<\Gamma^*(\cR,r)\), the sensitivity set implies \(V(\cR)\le r\).
Thus, larger values of \(\Gamma^*(\cR,r)\) indicate that stronger hidden bias
would be needed before the collective evidence for effects within \(\cR\)
can be overturned. For example, \(r=\lfloor |\cR|/2\rfloor\) corresponds to
assessing whether at least half of the outcomes in \(\cR\) are
affected.

\paragraph{Application 2: selecting robust subsets of outcomes.}

In an exploratory study with many outcomes, researchers may wish to identify
scientifically meaningful subsets for which the evidence that some fraction of the outcomes are affected remains robust to unmeasured confounding. The generalized sensitivity value provides a natural path forward.
Specifically, suppose a researcher considers subsets of size \(m\) among
\(K\) outcomes, yielding \(\binom{K}{m}\) candidate subsets. For a specified
tolerance \(r<m\) on the number of true null hypotheses, the researcher may
compare \(\Gamma^*(\cR,r)\) across the candidate subsets \(\cR\). For every
\(\Gamma<\Gamma^*(\cR,r)\), the sensitivity set supports the conclusion that
at least \(m-r\) outcomes in \(\cR\) are affected. Thus, larger values of
\(\Gamma^*(\cR,r)\) identify subsets for which the claim that at least \(m-r\) outcomes in \(\cR\) are affected
persists over a wider range of hidden bias. Because the sensitivity sets are
simultaneously valid over all \(\cR\), the reported confidence statement
remains valid even when the subset is chosen after comparing the candidate subsets. 
\subsection{A naive approach to controlling the false discovery proportion}\label{subsec:naive}
A seemingly natural approach to calculate $v_{\Gamma}^*(\cR)$ would be to first compute the worst-case \( p \)-values \(p_{k,\Gamma}^*\) for each hypothesis individually as in Equation~\eqref{eq:worst_pval}, and
substitute them for \(p_{k,\brho}\) in the Bonferroni local test in
Equation~\eqref{eq:bonfrho}. This gives $\phi_{\cI,\Gamma}^{\text{naive}} = \mathbf{1}\left\{ \min_{k \in \cI}\, p_{k,\Gamma}^* \leq \frac{\alpha}{|\cI|} \right\}.$
Then, the closed testing procedure based on these local tests is consonant, and Holm-Bonferroni can be used to determine which elementary hypotheses are rejected, 
and the resulting bound, which we denote by
\(v_{\Gamma}^{\mathrm{naive}}(\cR)\), can be computed for any
\(\cR\) using Equation~\eqref{eq:vR_cons}.

While this approach provides a valid FDP confidence set for any $\cR$, it is generally conservative, often dramatically so. This conservativeness arises because the worst-case \( p \)-values \( p_{k,\Gamma}^* \) are computed separately for each hypothesis \( k \), and the worst-case hidden bias configurations $\brho$ that maximize these \( p \)-values are allowed to vary across hypotheses. 
While $\brho$ is unknown, there is a common configuration of the hidden bias across the outcome variables.
To see the potential for improvements,  consider instead finding a single configuration \( \brho \) that most challenges the rejection of \( H_{\cI} \) though minimizing \( \phi_{\cI,\brho} \) over all allowable \( \brho \), resulting in
\begin{equation}\label{eq:SA_local}
   \phi_{\cI,\Gamma}^* := 
\min_{ \brho \in \mathcal{P}_\Gamma } \phi_{\cI,\brho} =  \mathbf{1}\left\{ \max_{ \brho \in \mathcal{P}_\Gamma } \;\min_{k \in \cI} \,p_{k,\brho} \leq \frac{\alpha}{|\cI|} \right\}. 
\end{equation}
This provides a uniform improvement on the $\phi_{\cI,\Gamma}^{\text{naive}}$ due to the minimax inequality \citep{fogarty2016sensitivity}:

\begin{equation}\label{eq:minimax}
    \phi_{\cI,\Gamma}^{\text{naive}} = \mathbf{1}\left\{ \min_{k \in \cI}\; \max_{ \brho \in \mathcal{P}_\Gamma } \,p_{k,\brho} \leq \frac{\alpha}{|\cI|} \right\}
 \leq \mathbf{1}\left\{ \max_{ \brho \in \mathcal{P}_\Gamma } \;\min_{k \in \cI} \,p_{k,\brho}\leq \frac{\alpha}{|\cI|} \right\} = \phi_{\cI,\Gamma}^*.
\end{equation}
Interestingly, as will be shown in Section~\ref{subsec:exact_SA}, exploiting the improved sensitivity analysis for each local test on the right hand side of Equation~\eqref{eq:SA_local} turns out to be sufficient to obtain the exact quantity $v_{\Gamma}^*(\cR)$. However, perhaps surprisingly, this comes with the price of moving away from a consonant closed testing procedure, even for Bonferroni-based procedures, as we will illustrate in Section~\ref{subsubsec:non_consonance}. 

\subsection{More powerful false discovery control}\label{subsec:exact_SA}
\subsubsection{Deriving $v_{\Gamma}^*(\cR)$ through closed sensitivity analysis}
\label{subsubsec:worst_case_fdp}

To improve power over the naive approach, we return to the exact worst-case FDP bound $v_{\Gamma}^*(\cR)
    :=
    \max_{\brho \in \mathcal{P}_\Gamma} v_{\brho}(\cR),$ suppressing its dependence on the testing level $\alpha$. By Equation~\eqref{eq:vR},
this quantity is the largest size of a subset $\cI\subseteq\cR$ whose intersection
hypothesis $H_{\cI}$ is not rejected by the closed testing procedure under some allowable $\brho \in \cP_\Gamma$. The following proposition shows that the local sensitivity tests in Equation~\eqref{eq:SA_local} are sufficient to identify this quantity.

\begin{proposition}
\label{prop:closed_sensitivity_representation}
For every $\cJ\subseteq[K]$, let
$\phi_{\cJ,\Gamma}^*$ be defined as in
\eqref{eq:SA_local}. Then, for every $\cR\subseteq[K]$,
\[
v_\Gamma^*(\cR)
=
\max_{\cJ\subseteq[K]}
|\cJ\cap\cR|
\left(
1-\phi_{\cJ,\Gamma}^*
\right)
\leq
\max_{\cJ\subseteq[K]}
|\cJ\cap\cR|
\left(
1-\phi_{\cJ,\Gamma}^{\mathrm{naive}}
\right)
=v_\Gamma^{\mathrm{naive}}(\cR).
\]
\end{proposition}

Thus, computing the local sensitivity tests
\(\phi_{\cJ,\Gamma}^*\) for all \(\cJ\subseteq[K]\) is sufficient to
obtain the exact worst-case FDP bound for every \(\cR\subseteq[K]\).
Moreover, Proposition~\ref{prop:closed_sensitivity_representation}
shows that this bound is never larger than the bound produced by the
naive approach. We refer to the closed testing procedure based on the
local sensitivity tests \(\phi_{\cJ,\Gamma}^*\) as a
\emph{closed sensitivity analysis}.

\subsubsection{Non-consonance in the closed sensitivity analysis}\label{subsubsec:non_consonance}
While the closed sensitivity analysis provides the exact sensitivity sets for the FDP, it necessitates moving away from a consonant closed testing procedure. That is, \emph{non-consonant rejections} are inevitable in the closed sensitivity analysis. A rejection of an intersection null hypothesis \( H_{\cI} \) is non-consonant if none of the elementary hypotheses \( H_k \) for \( k \in \cI \) are rejected.
Figure~\ref{fig:nonconsonance} (a), adapted from
\citet[Figure~1]{goemansolari}, illustrates such a rejection. Crosses
indicate hypotheses rejected by the closed sensitivity analysis. In
particular, \(H_2\cap H_3\) is rejected, whereas neither \(H_2\) nor
\(H_3\) is rejected.
To see how this can arise, consider the local Bonferroni test of
\(H_2\cap H_3\). Its sensitivity analysis rejects when $\min\{p_{2,\brho},p_{3,\brho}\}
\le {\alpha}/{2}$ for every $\brho\in\cP_\Gamma.$ Equivalently, the two hatched regions in
Figure~\ref{fig:nonconsonance}(b) cover \(\cP_\Gamma\).
At the same time, the singleton sensitivity tests can fail to reject if
there exist, possibly different, configurations
\(\brho_2,\brho_3\in\cP_\Gamma\) such that $p_{2,\brho_2}>\alpha$ and $p_{3,\brho_3}>\alpha.$
Thus, different hidden bias configurations can prevent rejection of
\(H_2\) and \(H_3\) separately, even though no single configuration
prevents rejection of their intersection. See Figure~\ref{fig:nonconsonance}(b) for this visualization.

Such non-consonant rejections are impossible in the conventional Bonferroni-based closed testing procedure for a particular $\brho$, which is consonant by design. 
However, this phenomenon is precisely what allows the closed sensitivity analysis to
produce sharper FDP sensitivity sets. As noted by \cite{goemansolari}, non-consonant rejections can yield sharper confidence sets beyond what can be obtained from elementary rejections alone. To describe this gain, write $\widetilde{\phi}_{\cI,\Gamma}^*
=
\min_{\cJ\supseteq\cI}\phi_{\cJ,\Gamma}^*$, fix a set 
\(\cR\subseteq[K]\), and consider $U_\Gamma(\cR)
    =
    \left\{
    k\in\cR:
    \tilde\phi_{\{k\},\Gamma}^*=0
    \right\}.$
This is the set of elementary hypotheses in \(\cR\) that are not rejected by the
closed sensitivity analysis. If the closed sensitivity procedure were consonant, by Equation~\eqref{eq:vR_cons}, the FDP bound would be the number of unrejected elementary hypotheses,
namely $v_\Gamma^*(\cR)=|U_\Gamma(\cR)|.$
Thus \(|U_\Gamma(\cR)|\) is the FDP upper bound that would be obtained from
elementary rejections alone. Non-consonant rejections can improve upon this bound, as we formalize in the following.

\begin{proposition}
\label{prop:fdp_gain_nonconsonance}
$v_\Gamma^*(\cR)\le |U_\Gamma(\cR)|.$
Moreover, $v_\Gamma^*(\cR)<|U_\Gamma(\cR)|$
if and only if the closed sensitivity analysis rejects the intersection null for \(U_\Gamma(\cR)\). In this case, the rejection is non-consonant.
\end{proposition}
For the example in Figure~\ref{fig:nonconsonance}(a), take
\(\cR=\{2,3\}\). Since neither elementary hypothesis is rejected,
elementary rejections alone give the upper bound
\(|U_\Gamma(\cR)|=2\) on the number of true nulls in \(\cR\),
and hence the conservative FDP sensitivity set \(\{0,1/2,1\}\). The closed sensitivity analysis nevertheless rejects
\(H_{U_\Gamma(\cR)}=H_2\cap H_3\).
Proposition~\ref{prop:fdp_gain_nonconsonance} therefore gives
\(v_\Gamma^*(\cR)<2\). Because each singleton remains unrejected,
\(v_\Gamma^*(\cR)\ge1\), and hence $v_\Gamma^*(\cR)=1.$
The resulting FDP sensitivity set is \(\{0,1/2\}\): the analysis
certifies that at least one of \(H_2\) and \(H_3\) is false, although it
cannot identify which one.
This is precisely the additional information extracted from the non-consonant rejection of \(H_2\cap H_3\).

\begin{figure}[t]
\centering

\begin{minipage}[c]{0.42\textwidth}
    \centering
\pgfdeclarelayer{bg}
\pgfsetlayers{bg,main}
\begin{tikzpicture}[
    scale =0.8,
    node distance=1cm and 1cm,
    node style/.style={
        rectangle,
        line width=0.7mm,
        draw,
        minimum width=1.2cm,
        minimum height=0.4cm,
        align=center
    }
]

\node[node style] (H1H2H3) at (0,0) {$H_1 \cap H_2 \cap H_3$};
\node[node style] (H1H2) at (-3,-2) {$H_1 \cap H_2$};
\node[node style] (H1H3) at (0,-2) {$H_1 \cap H_3$};
\node[node style] (H2H3) at (3,-2) {$H_2 \cap H_3$};
\node[node style] (H1) at (-3,-4) {$H_1$};
\node[node style] (H2) at (0,-4) {$H_2$};
\node[node style] (H3) at (3,-4) {$H_3$};

\begin{pgfonlayer}{bg}
    \foreach \n in {H1H2H3, H1H2, H1H3, H1} {
        \draw[black, line width=0.5mm]
            (\n.south west) -- (\n.north east)
            (\n.south east) -- (\n.north west);
    }

    \draw[black, line width=0.5mm, dashed]
        (H2H3.south west) -- (H2H3.north east)
        (H2H3.south east) -- (H2H3.north west);
\end{pgfonlayer}

\draw[->, line width=0.5mm] (H1H2H3) -- (H1H2);
\draw[->, line width=0.5mm] (H1H2H3) -- (H1H3);
\draw[->, line width=0.5mm] (H1H2H3) -- (H2H3);

\draw[->, line width=0.5mm] (H1H2) -- (H1);
\draw[->, line width=0.5mm] (H1H2) -- (H2);

\draw[->, line width=0.5mm] (H1H3) -- (H1);
\draw[->, line width=0.5mm] (H1H3) -- (H3);

\draw[->, line width=0.5mm] (H2H3) -- (H2);
\draw[->, line width=0.5mm] (H2H3) -- (H3);

\end{tikzpicture}\caption*{(a) Closed testing tree.}
\end{minipage}
\hfill
\begin{minipage}[c]{0.56\textwidth}
    \centering

\begin{tikzpicture}[
    x=0.78cm,
    y=0.78cm,
    line cap=round,
    line join=round
]

\begin{scope}
    \clip (-4.5,-1.8) rectangle (4.5,1.8);

    \fill[
        pattern=vertical lines,
        pattern color=black!55
    ]
    (-4.5,-1.8)
    -- (0.35,-1.8)
    .. controls (0.65,-0.9) and (0.25,0.9) .. (0.50,1.8)
    -- (-4.5,1.8)
    -- cycle;

    \fill[
        pattern=horizontal lines,
        pattern color=black!55
    ]
    (4.5,-1.8)
    -- (-0.35,-1.8)
    .. controls (-0.65,-0.9) and (-0.25,0.9) .. (-0.50,1.8)
    -- (4.5,1.8)
    -- cycle;

    \draw[thick]
    (0.35,-1.8)
    .. controls (0.65,-0.9) and (0.25,0.9) .. (0.50,1.8);

    \draw[thick]
    (-0.35,-1.8)
    .. controls (-0.65,-0.9) and (-0.25,0.9) .. (-0.50,1.8);
\end{scope}

\draw[thick] (-4.5,-1.8) rectangle (4.5,1.8);
\node[anchor=south west] at (-4.5,1.82) {$\cP_\Gamma$};

\node[
    font=\scriptsize,
    fill=white,
    inner sep=1.5pt
] at (-2.35,0.85)
{$\left\{\brho:p_{2,\brho}\le\alpha/2\right\}$};

\node[
    font=\scriptsize,
    fill=white,
    inner sep=1.5pt
] at (2.35,0.85)
{$\left\{\brho:p_{3,\brho}\le\alpha/2\right\}$};

\fill (-2.45,-0.45) circle (1.8pt);
\node[below=2pt, align=center, font=\scriptsize]
at (-2.45,-0.45)
{
    $\brho_3$\\[-1pt]
    $p_{3,\brho_3}>\alpha$
};

\fill (2.45,-0.45) circle (1.8pt);
\node[below=2pt, align=center, font=\scriptsize]
at (2.45,-0.45)
{
    $\brho_2$\\[-1pt]
    $p_{2,\brho_2}>\alpha$
};

\end{tikzpicture}\caption*{(b) Hidden bias configurations.}
    \label{fig:nonconsonance-configurations}
\end{minipage}

\caption{\small
A non-consonant rejection in the closed sensitivity analysis.
Panel~(a), adapted from \citet[Figure~1]{goemansolari}, shows the
closed testing tree. Crosses indicate hypotheses rejected; the dashed cross highlights the non-consonant rejection of
\(H_2\cap H_3\).
In panel~(b), vertical and horizontal hatching indicate configurations
satisfying \(p_{2,\brho}\le\alpha/2\) and
\(p_{3,\brho}\le\alpha/2\), respectively. These regions cover
\(\cP_\Gamma\), so the local test of \(H_2\cap H_3\) rejects, whereas
\(\brho_2\) and \(\brho_3\) provide failures to reject \(H_2\) and
\(H_3\), respectively.
}
\label{fig:nonconsonance}

\end{figure}

In the presence of non-consonant rejections, we therefore cannot rely solely on elementary rejections to determine \(v_\Gamma^*(\cR)\). Instead, we must examine higher-level intersection hypotheses within the closed testing tree. Computing these intersection tests
efficiently is therefore central to obtaining the sharper bound
\(v_\Gamma^*(\cR)\), motivating the methods developed in the next
section.

\section{Computation for matched observational studies}
\label{sec:computation}
The methodology in Section~\ref{sec:SA_FDP} applies to any sensitivity model
that can be represented by a set \(\cP_\Gamma\) of hidden bias
configurations. We now describe
the computation of \(v_\Gamma^*(\cR)\) under Rosenbaum's sensitivity model in
Example~\ref{ex:rosenbaum}. Of the three examples considered in Section \ref{subsec:sensitivity_model}, multiple testing methodology for sensitivity analysis is most developed under Rosenbaum's sensitivity model \citep{ros16, fogarty2016sensitivity}. However, as forthcoming developments illustrate, a naive application of existing methods does not yield a computationally scalable framework for FDP control due to the number and size of the required optimization problems. Instead, to facilitate sensitivity analysis for FDP control in settings with many outcome variables, we develop a new optimization routine directly targeting $v_\Gamma^*(\mathcal{R})$.

\subsection{Test statistics and null distributions}
For notational convenience, we assume that within each matched set $\mcm_b$, there is one treated unit and \(  |\mcm_b|-1 \geq 1 \) control units. Extensions to full matching, where multiple treated units are allowed, are immediate; See \citet[][Chapter 4, Problem 12]{Rosenbaum2002}.
Let \(\Omega=\{\bz\in\{0,1\}^N:\sum_{i\in\mathcal M_b}z_i=1,\ b=1,\ldots,B\}\), then \(\cZ\) in Example~\ref{ex:rosenbaum} is the event \(\bZ\in\Omega\), and \( |\Omega| = \prod_{b=1}^B |\mcm_b| \). For a fixed \(\brho\in\cP_\Gamma\), treatment assignments are independent across matched sets and

\[
    \P_{\brho}(\bZ=\bz\mid \cF,\cZ)
    =
    \prod_{b=1}^B\prod_{i\in\mathcal M_b}\varrho_i^{z_i},
    \qquad \bz\in\Omega.
\]

For each outcome \( k \), consider Fisher's sharp null hypothesis of no causal effect:

\begin{singlespace}
\begin{equation}\label{eq:sharpnull}
    H_k: r_{Tik} = r_{Cik}\;\;\;\forall i.
\end{equation}
\end{singlespace}
\noindent
Extensions to other types of hypotheses can be obtained; see \citet[][Section 5]{Rosenbaum2002} and \citet[][Sections 2.4-2.5]{designofobs} for an overview. 
We use
test statistics of the form

\begin{singlespace}
\begin{equation}
\label{eq:matched_test_stat}
    T_k(\bZ)
    =
    \bZ^\top\bq_k
    =
    \sum_{b=1}^B\sum_{i\in\mathcal M_b} Z_i q_{ik},
\end{equation}
\end{singlespace}
\noindent
where $\bq_k=(q_{1k},\ldots,q_{Nk})^\top$ and $q_{ik}$ is some function of the observed outcomes $\bR_k$. 
This form of test statistic covers a wide range of test statistics as special cases, see \citet{Rosenbaum2002} for more details.
Let \(t_k=T_k(\bz^{\rm obs})\) denote the observed value.
Under $H_k$, where $\bR_{k} = \br_{Ck}$, $\bq_k$ becomes a function of $\br_{Ck}$, and is thus a fixed quantity across all $\bz \in \Omega$. Therefore, under a fixed $\brho$, the distribution of \(T_k(\bZ)\) is

\[
    \P_{\brho}\{T_k(\bZ)\ge a\mid\cF,\cZ;H_k\}
    =
    \sum_{\bz\in\Omega}
    \mathbf 1\{T_k(\bz)\ge a\}
    \prod_{b=1}^B\prod_{i\in\mathcal M_b}\varrho_i^{z_i}.
\]
Therefore, a single $\brho$ determines the null distributions of all $K$ test statistics simultaneously.

\subsection{Normal approximation and asymptotic validity}
For a fixed \(\brho\in\cP_\Gamma\), let
\(\mu_{k,\brho}=\E_{\brho}(T_k\mid\cF,\cZ;H_k)=\sum_{b=1}^B\sum_{i\in\mcm_b}\varrho_iq_{ik}\) and
\(\sigma_{k,\brho}^2=\Var_{\brho}(T_k\mid\cF,\cZ;H_k)=\sum_{b=1}^B
    \left\{
    \sum_{i\in\mcm_b}\varrho_iq_{ik}^2
    -
    \left(\sum_{i\in\mcm_b}\varrho_iq_{ik}\right)^2
    \right\}.\) 
Under suitable regularity conditions,  we have the asymptotic normality for each $T_k$:
\begin{equation}\label{eq:t-normal}
    \frac{T_k-\mu_{k,\brho}}{\sigma_{k,\brho}}
    \xrightarrow{\;d\;}
    \mathcal N(0,1),
\end{equation}
see Section \ref{app:comp_proofs} of the Supplementary Material for a description of sufficient conditions for this convergence in distribution.

Assuming that the elementary hypotheses $H_k$ for all $k \in [K]$ have two-sided alternatives, define $\zeta_k(\brho;c)
    :=
    (t_k-\mu_{k,\brho})^2
    -
    \chi^2_{1,1-c}\sigma_{k,\brho}^2,$
where \(\chi^2_{1,1-c}\) is the \((1-c)\)-quantile of a
\(\chi^2_1\) distribution. Under \eqref{eq:t-normal}, the event
\(p_{k,\brho}\leq c\) is approximated by
\(\zeta_k(\brho;c)\geq0\). Consequently, the local sensitivity test in
\eqref{eq:SA_local} can be approximated by $\phi_{\cJ,\Gamma}^* = \1\set{\min_{\adverrho}\underset{k \in \cJ}{\max}\;\zeta_k(\brho;\alpha/|\cJ|) \geq 0}.$ By Proposition~\ref{prop:closed_sensitivity_representation}, we may further write

\begin{equation}\label{eq:v_optimization}
v_\Gamma^*(\cR)
=
\max_{\cJ\subseteq[K]}
|\cJ\cap\cR|
\,
\1\left\{
\min_{\brho\in\cP_\Gamma}
\max_{k\in\cJ}
\zeta_k\left(\brho;\frac{\alpha}{|\cJ|}\right)
<0
\right\}.
\end{equation}
The following theorem provides conditions under which the FDP bounds obtained using the normal approximation retain asymptotic validity.

\begin{theorem}[Asymptotic simultaneous validity]
\label{thm:normal_validity}
Let \(\brho_0\) denote the true hidden bias configuration, and suppose that
\(\brho_0\in\cP_{\Gamma_0}\). Assume that the regularity condition  $\max_{k\in[K]}\sup_{\brho\in\cP_\Gamma}$ $\sum_{b=1}^B$ $
\E_{\brho}\!\left[
\left|
\sum_{i\in\mathcal M_b}(Z_i-\varrho_i)q_{ik}
\right|^3
\,\middle|\,\cF,\cZ
\right]
/{
\sigma_{k,\brho}^3
}
\longrightarrow 0$ holds. Then the FDP bounds based on normally
approximated two-sided \(p\)-values and Bonferroni local tests satisfy
\[
\P_{\brho_0}\!\left\{
V(\cR)\leq v_\Gamma^*(\cR)
\ \text{for every }\cR\subseteq[K]
\
\right\}
\geq 1-\alpha-o(1)
\]
for every $\Gamma\geq\Gamma_0$.
Consequently, $\left\{
0,\frac{1}{|\cR|},\ldots,
\frac{v_\Gamma^*(\cR)}{|\cR|}
\right\}$
defines an asymptotically simultaneous \(100(1-\alpha)\%\)
sensitivity set for the FDP over all nonempty
\(\cR\subseteq[K]\), for every \(\Gamma\geq\Gamma_0\).
\end{theorem}

\subsection{Computing \(v_\Gamma^*(\cR)\) through integer programming}
Equation~\eqref{eq:v_optimization} shows that
\(v_\Gamma^{*}(\cR)\) is the largest overlap of \(\cR\) with a set \(\cJ\) whose local sensitivity test fails to reject \(H_\cJ\). 
For any fixed \(\cJ\), the corresponding local sensitivity test $\phi_{\cJ,\Gamma}^*$ can be
computed using the method of \citet{fogarty2016sensitivity}. Thus,
\eqref{eq:v_optimization} provides a direct approach to computing
\(v_\Gamma^*(\cR)\): apply their method separately to every relevant set
\(\cJ\). This approach, however, requires enumerating
exponentially many sets in the worst case, and is computationally prohibitive.
Rather than solving a separate optimization problem for each relevant
\(\cJ\), we use binary decision variables to select $\cJ$ so that one optimization
problem searches simultaneously over all sets \(\cJ\) having a
prescribed overlap \(r=|\cJ\cap\cR|\).
Because \(v_\Gamma^*(\cR)\) is the largest such overlap, we consider
\(r=|\cR|,|\cR|-1,\ldots,1\) and stop at the first value for which a
qualifying set \(\cJ\) is found.
For a fixed \(r\), the condition in \eqref{eq:v_optimization} also
depends on \(v=|\cJ|\) through the Bonferroni threshold. Since \(\cJ\)
contains \(r\) elements of \(\cR\) and at most all \(K-|\cR|\) elements
outside \(\cR\), it suffices to consider $v=r,\ldots,r+K-|\cR|.$
For each pair \((r,v)\), introduce binary variables
\(\theta_k\in\{0,1\}\), where \(\theta_k=1\) indicates that
\(k\in\cJ\). The constraints $\sum_{k\in\cR}\theta_k=r$ and $\sum_{k=1}^K\theta_k=v$
allow the program to search over all sets \(\cJ\) having the prescribed
overlap and cardinality. 
The auxiliary variable \(y\)
represents the largest value of
\(\zeta_k(\brho;\alpha/v)\) among the selected outcomes. Minimizing \(y\)
over both \(\cJ\) and \(\brho\in\cP_\Gamma\) therefore determines whether
there exists such a \(\cJ\) for which $\min_{\brho\in\cP_\Gamma}
\max_{k\in\cJ}
\zeta_k\left(\brho;{\alpha}/{v}\right)<0.$ The resulting procedure is given in
Algorithm~\ref{alg:IP}. For other uses of integer programming in the design and analysis of
matched observational studies, see
\citet{zub12,pimentel2015large,zhang2024sensitivity,heng2025sensitivity}.

\begin{singlespace}
\begin{algorithm}[h!]
\caption{Integer program for computing \(v_\Gamma^*(\cR)\)}
\label{alg:IP}
\begin{mdframed}
\textbf{For} \(r=|\cR|,|\cR|-1,\ldots,1\):\\
\quad \textbf{For} \(v=r,\ldots,r+K-|\cR|\):
\begin{align*}
    \min_{y,\boldsymbol\theta,\brho}\quad &y\\
    \text{s.t.}\quad
    &\sum_{k\in\cR}\theta_k=r,
    \qquad
    \sum_{k=1}^K\theta_k=v,\\
    &y\geq
    \zeta_k\left(\brho;{\alpha}/{v}\right)
    \quad\text{if }\theta_k=1,
    \qquad k=1,\ldots,K,\\
    &y<0,\\
    &\theta_k\in\{0,1\},
    \qquad
    \brho\in\cP_\Gamma.
\end{align*}
\quad \textbf{If} the problem is feasible, \textbf{return} \(r\).\\
\textbf{Return} \(0\).
\end{mdframed}
\end{algorithm}
\end{singlespace}

\subsection{Screening rules}\label{sec:screening}

Algorithm~\ref{alg:IP} searches over possible values of \(r\) and \(v\).
To reduce the number of integer programs that must be solved, we first apply
screening rules based on the individual worst-case \(p\)-values
\(p_{k,\Gamma}^*\) in \eqref{eq:worst_pval}. These worst-case \(p\)-values
can be computed under the normal approximation without integer programming;
see Section~\ref{app:details} of the Supplementary Material.

To provide intuition for these rules,  consider a singleton
\(\cR=\{k\}\), then
\(v_\Gamma^*(\cR)=1-\tilde\phi_{\{k\},\Gamma}^*\), so it suffices to
determine whether \(H_k\) is rejected by the closed sensitivity analysis. If
\(p_{k,\Gamma}^*\leq\alpha/K\), then every local sensitivity test containing
\(H_k\) rejects, and hence the closed sensitivity analysis rejects \(H_k\).
If \(p_{k,\Gamma}^*>\alpha\), then the local sensitivity test for \(H_k\)
itself fails to reject, and hence the closed sensitivity analysis also fails
to reject \(H_k\). This gives the screening rule
\begin{singlespace}
    \begin{equation}\label{eq:screening}
\text{Decision for }H_k=
\begin{cases}
\text{reject }H_k,
& p_{k,\Gamma}^*\leq\alpha/K,\\
\text{fail to reject }H_k,
& p_{k,\Gamma}^*>\alpha,\\
\text{proceed with Algorithm~\ref{alg:IP}},
& \alpha/K<p_{k,\Gamma}^*\leq\alpha.
\end{cases}
\end{equation}
\end{singlespace}

\noindent
Thus, the integer program is needed only when the worst-case \(p\)-value lies
between \(\alpha/K\) and \(\alpha\). The corresponding screening rules for a
general set \(\cR\), which eliminate candidate values of \(r\) and \(v\),
are provided in Section~\ref{app:comp} of the Supplementary Material.

The following theorem highlights the merits of screening using individual $p$-values. It shows that under mild regularity conditions, this screening step alone suffices asymptotically to determine the sensitivity set for the FDP when considering singleton sets $\mathcal{R}$.

\begin{theorem}\label{thm:IP}
Let
\(\mu_k=\E(T_k\mid\cF,\cZ)\) be the actual expectation of $T_k$ and suppose that $\max_{k\in[K]}{|T_k-\mu_k|}/{B}\overset{p}{\longrightarrow}0$, and
let \(\mu_{k,\Gamma}^*\) and \(\sigma_{k,\Gamma}^*\) be the expectation and standard deviation under a configuration attaining the worst-case $p$-value
\(p_{k,\Gamma}^*\). Suppose that $\limsup_{B\to\infty}
\max_{k\in[K]}
\frac{1}{B}\sum_{b=1}^B \max_{i\in\mathcal M_b}|q_{ik}|^2
<\infty.$
Suppose further that there exists \(\epsilon>0\) such that, for every \(k\in[K]\), $\P\left(\mu_{k,\Gamma}^*=T_k\;\text{or}\;
\frac{|\mu_k-\mu_{k,\Gamma}^*|}{B}\ge\epsilon
\right)
\longrightarrow1$.
Then $\P\left(
\bigcup_{k=1}^K
\left\{
\frac{\alpha}{K}<p_{k,\Gamma}^*\le\alpha
\right\}
\right)
\longrightarrow0.$
Consequently, the probability of invoking the integer program when
evaluating all singleton sets converges to zero.
\end{theorem}

\begin{remark}\label{rem:IP_growingK}
The conclusion of Theorem~\ref{thm:IP} continues to hold when $K$ grows with $B$, under suitable conditions. See Section~\ref{app:proof_growingK} of the Supplementary Material for more details.
\end{remark}

For details of the screening rules for general $\cR$ to reduce the search of the values for both $r$ and $v$ in the integer program, see Section ~\ref{app:comp} of the Supplementary Material. 
We provide an implementation for Algorithm~\ref{alg:IP} and the screening rules in Python using Gurobi. 
In Section~\ref{sec:simu_runtime} and Section~\ref{sec:simu_screening}, we further provide simulation studies to  demonstrate the efficiency of our integer program and the screening rules.


\section{Simulation study}\label{sec:simulation}

We evaluate the proposed methods under the setting of Example~\ref{ex:rosenbaum}, specializing to matched pairs.
Thus, the \(N=2B\) units are partitioned into
\(\mathcal M_1,\ldots,\mathcal M_B\), with
\(|\mathcal M_b|=2\) and exactly one treated unit in each matched pair.
Throughout, the testing level is \(\alpha=0.05\).
\subsection{Simulation setup}
Let \(B\) denote the number of matched pairs and \(K\) the number of outcomes.
For each unit \(i\in[N]\), generate the \(K\)-dimensional potential outcomes through 
\begin{equation}
\label{eq:simu_setting}
    \br_{Ci}
    \stackrel{\mathrm{iid}}{\sim}
    \mathcal N(\mathbf 0_K,\Sigma),
    \qquad
    \br_{Ti}
    =
    \br_{Ci}+\btau,
\end{equation}
where \(\Sigma\) is the covariance matrix of the outcomes and
\(\btau=(\tau_1,\ldots,\tau_K)^\top\) is the treatment-effect vector.
The sharp null hypothesis \(H_k\) in Equation~\eqref{eq:sharpnull} is true if and only if \(\tau_k=0\).

For the first three simulation studies, we consider the following settings. Two treatment-effect vectors are used:
\begin{equation}
\label{eq:btau}
    \tau_k^{(1)}
    =
    0.15+\frac{0.20(k-1)}{K-1},
    \qquad
    \tau_k^{(2)}
    =
    0.3\,\mathbf 1\{k\le K/2\},
    \qquad k\in[K].
\end{equation}
Thus, the components of \(\btau^{(1)}\) increase linearly from \(0.15\)
to \(0.35\), whereas the first \(K/2\) components of \(\btau^{(2)}\)
equal \(0.3\) and the remaining components equal zero.
We use the covariance matrices
\begin{equation}
\label{eq:Sigma_matrices}
    \Sigma^{(1)}=\mathbf I_K,
    \qquad
    \Sigma^{(2)}
    =
    (1-\eta)\mathbf I_K
    +
    \eta\mathbf 1_K\mathbf 1_K^\top,
    \qquad
    \eta=0.2.
\end{equation}
Under \(\Sigma^{(1)}\), the outcomes are independent, whereas
\(\Sigma^{(2)}\) has equicorrelation \(0.2\).
Treatment is assigned uniformly within each matched pair, corresponding to NUC setting. 
We nevertheless conduct sensitivity analyses at values \(\Gamma>1\) to evaluate robustness to departures from the NUC setting. For each outcome, we construct the statistic \(T_k\)
in Equation~\eqref{eq:matched_test_stat} using Huber's M-scores with outer
trimming constant \(2.5\) \citep{hub81,ros07}.
\subsection{Improved power over the naive approach}
\begin{figure}[h]
    \centering
    \includegraphics[width=0.92\textwidth]{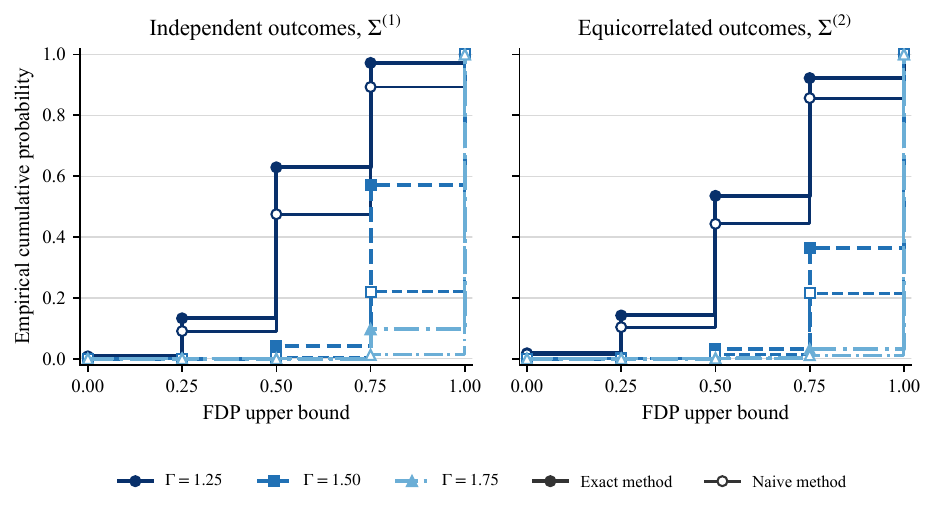}
    \caption{\small Empirical distributions of the FDP upper bounds from
    the exact and naive methods. A curve that lies higher corresponds to a
    stochastically smaller, and hence tighter, bound. 
    Colours, line types, and marker shapes distinguish the values of $\Gamma$, while filled and open markers represent the exact and naive methods, respectively. Results for $\Gamma=1$ are omitted because the two methods coincide.
    }
    \label{fig:bound_distributions}
\end{figure}

This section demonstrates the power of our closed sensitivity analysis in Section~\ref{subsec:exact_SA} with the naive approach in Section~\ref{subsec:naive} through comparing the FDP upper bound obtained by the two approaches. 
We set \(B = 500\) and consider \(K = 4\) outcomes. 
We use $\btau^{(1)}$ in Equation~\eqref{eq:btau} as the treatment-effect vector and use both $\Sigma^{(1)}$ and $\Sigma^{(2)}$ in Equation~\eqref{eq:Sigma_matrices} to simulate data. For each dataset, we apply both the exact and naive approaches and compute
\(v_{\Gamma}^*(\cR)\) and \(v_{\Gamma}^{\mathrm{naive}}(\cR)\),
respectively, for \(\cR=[K]\) and
\(\Gamma\in\{1,1.25,1.5,1.75\}\). The corresponding FDP upper bounds are
\(v_{\Gamma}^*(\cR)/4\) and
\(v_{\Gamma}^{\mathrm{naive}}(\cR)/4\), both taking values in
\(\{0,1/4,\ldots,1\}\). Figure~\ref{fig:bound_distributions} compares
their empirical distributions over the 1,000 simulated datasets.

When $\Gamma = 1$, where no unmeasured confounding exists, the two methods are equivalent.
Figure~\ref{fig:bound_distributions} shows that, for \(\Gamma>1\), the bound from our method is stochastically no larger than the naive bound, indicating that the sensitivity intervals for the false discovery proportion produced by the naive method are wider, thereby reflecting the power gains of our proposed approach discussed in Section~\ref{subsec:naive}.
For example, when \(\Gamma=1.5\) and \(\Sigma=\Sigma^{(1)}\), the probability of obtaining an FDP upper bound at most \(0.75\) is 0.571 under our method and 0.221 under the naive method. Equivalently, our method returns the vacuous upper bound one in 42.9\% of simulations, compared with 77.9\% for the naive method. 


\subsection{Runtime comparison}\label{sec:simu_runtime}
We now compare the computational efficiency of Algorithm~\ref{alg:IP} with an enumerative approach based on \citet{fogarty2016sensitivity}. The two algorithms use the same screening rules and return the same sensitivity sets. When screening is inconclusive, however, the enumerative method evaluates the required intersection hypotheses one at a time, whereas Algorithm~\ref{alg:IP} considers all candidate intersections of a given cardinality in one optimization.

We set \(B = 500\) and \(K \in \{10, 20\}\). For each \(K\), we use  $\btau^{(1)}$ and $\btau^{(2)}$ in Equation~\eqref{eq:btau}
and both $\Sigma^{(1)}$ and $\Sigma^{(2)}$ in Equation~\eqref{eq:Sigma_matrices} to simulate data. 
For each pair of $(\btau, \Sigma)$, we generate 1,000 datasets and conduct the closed sensitivity analyses for all outcomes under $\Gamma \in \{1.25, 1.5, 1.75\}$. This yields 12 simulation settings in total for each \(K\), as arranged in Table~\ref{tab:12settings}.

\begin{singlespace}
    \begin{table}[htbp]
\centering
\begin{tabular}{c*{12}{c}}
\toprule
\(\boldsymbol{\tau}\) 
& \multicolumn{6}{c}{$\boldsymbol{\tau}^{(1)}$} 
& \multicolumn{6}{c}{$\boldsymbol{\tau}^{(2)}$} \\
\midrule

\(\Sigma\)
& \multicolumn{3}{c}{$\Sigma^{(1)}$}
& \multicolumn{3}{c}{$\Sigma^{(2)}$}
& \multicolumn{3}{c}{$\Sigma^{(1)}$}
& \multicolumn{3}{c}{$\Sigma^{(2)}$} \\
\midrule

\(\Gamma\)
& 1.25 & 1.5 & 1.75
& 1.25 & 1.5 & 1.75
& 1.25 & 1.5 & 1.75
& 1.25 & 1.5 & 1.75 \\
\midrule

\text{Setting}
& 1 & 2 & 3
& 4 & 5 & 6
& 7 & 8 & 9
& 10 & 11 & 12 \\
\bottomrule
\end{tabular}
\caption{The 12 simulation settings.}
\label{tab:12settings}
\end{table}
\end{singlespace}

\begin{figure}[h]
    \centering
\includegraphics[width=0.9\textwidth]{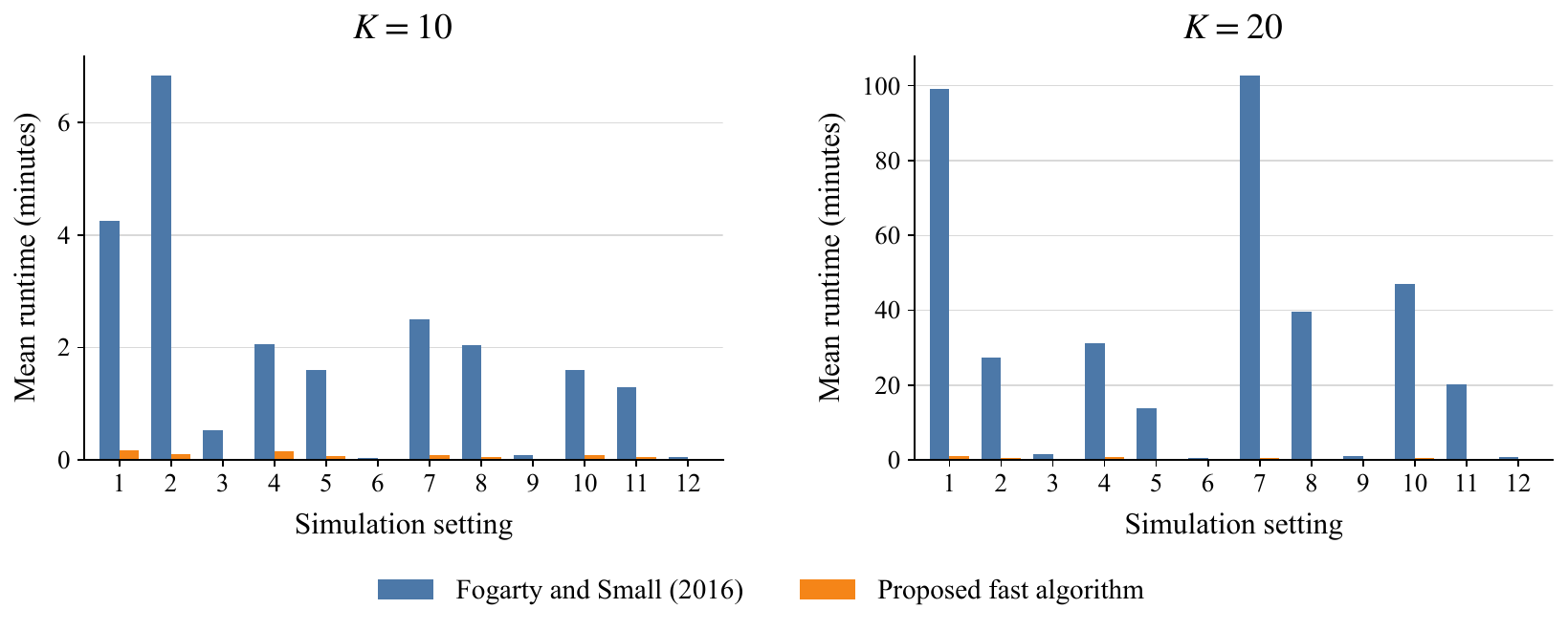}\caption{\small Runtime comparison (in minutes) of 12 settings in Table~\ref{tab:12settings}.}\label{fig:runtime_comparison}
\end{figure}

For each setting, we record the runtime needed to perform closed sensitivity analyses for all outcomes using both our proposal and the enumerative approach. To reiterate, we apply our screening rules to both approaches. The approaches thus differ solely in the efficiency with which they evaluate the components of the closed testing tree which screening cannot account for.  Figure~\ref{fig:runtime_comparison} compares the average runtimes. 
Across all the simulation settings, we observe reductions in computation time from our proposal ranging from 5-fold to 300-fold. 
Notably, in settings~1 and~7 with \(K = 20\) and \(B = 500\), enumerative closed testing takes more than one hour to complete, whereas our proposal finishes in only a few minutes.

\vspace{-0.8cm}
\subsection{Effectiveness of screening rules}\label{sec:simu_screening}
In this section we explore the benefits of the screening rules developed in~\eqref{eq:screening} for varying sample sizes, as considered by Theorem~\ref{thm:IP}. We set $K = 10$ and consider increasing sample sizes \(B \in \{500, 1000, 2000, 5000, 10000\}\). We use $\btau^{(2)}$ and \(\Sigma^{(1)}\) to generate the data.
For each $B$, we generate 1{,}000 datasets.
For each dataset, we perform the closed sensitivity analysis for each outcome using \(\Gamma \in \{1.25, 1.5, 1.75, 2.0\}\) and record (i) whether the integer program is invoked at least once and (ii) the fraction of the $K$ outcomes for which it is invoked. 

\begin{figure}[htbp]
    \centering
    \includegraphics[width=0.94\textwidth]{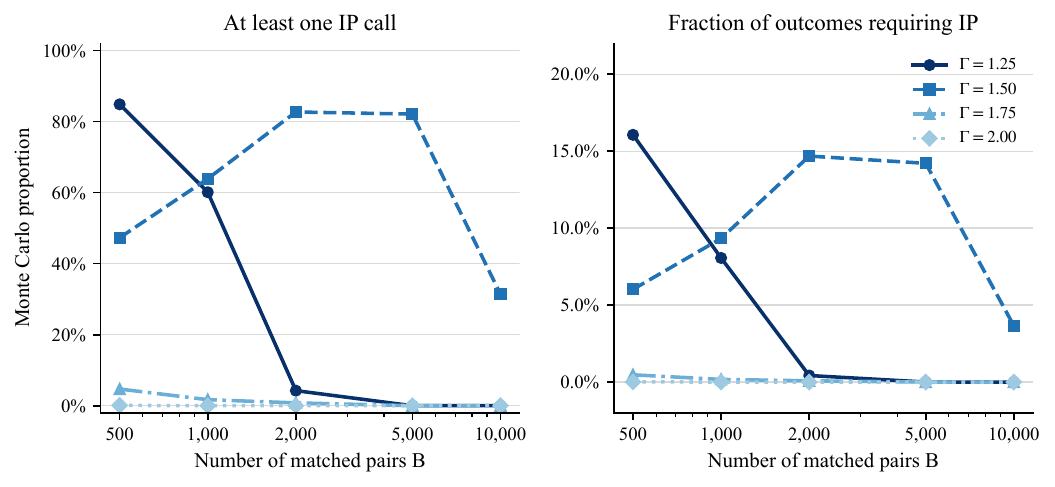}
    \caption{\small Performance of the screening rules as the number of matched pairs $B$ increases. }
    \label{fig:screening_simu}
\end{figure}

Figure~\ref{fig:screening_simu} presents the results. For
\(\Gamma=1.25,1.75,\) and \(2.00\), both the proportion of simulated
datasets requiring at least one integer program and the average fraction of
outcomes requiring integer programming decrease as \(B\) increases, in line
with Theorem~\ref{thm:IP}. For \(\Gamma=1.50\), the patterns are not monotone, as explained below. Overall,
the screening rules resolve most outcomes without integer programming,
thereby improving computational efficiency.

Specifically, note that an integer program is needed only when \(0.005<p_{k,\Gamma}^*\leq0.05\).
For \(\Gamma=1.25\), the worst-case \(p\)-values of the affected
outcomes increasingly fall below \(0.005\), so these outcomes are
rejected directly, whereas the null outcomes are generally screened
in the opposite direction because their worst-case \(p\)-values exceed
\(0.05\).
For \(\Gamma=1.75\) and \(2.00\), the worst-case \(p\)-values increasingly exceed \(0.05\).
For \(\Gamma=1.50\), the worst-case \(p\)-values lie near the two screening cutoffs, so sampling variation can affect whether an integer program is needed. As \(B\) grows and sampling variation decreases, more \(p\)-values first enter the intermediate interval \((0.005,0.05]\), increasing the probability that at least one integer program is needed from 47.3\% at \(B=500\) to 82.7\% at \(B=2000\). As \(B\) grows further, more \(p\)-values exceed \(0.05\), and this probability falls to 31.4\% at \(B=10{,}000\). Even at \(B=2000\), only 14.7\% of the hypotheses require integer programming on average, because most are resolved directly by screening.

\subsection{Selecting robust subsets of outcomes}
\label{subsec:gsv_subset_selection}
We now demonstrate how the generalized sensitivity value could be used to identify promising subsets of outcomes. We consider \(K=4\) outcomes, of which the first
two are affected by treatment and the last two satisfy their sharp null
hypotheses. Treatment assignment depends on the potential outcomes for
outcomes 3 and 4 so that individuals with higher values on these outcomes are more likely to be treated, causing these outcomes to exhibit spurious positive treated-minus-control differences.

A researcher, unaware of this, aims to select a subset with two outcomes for subsequent investigation that contains at least one outcome genuinely affected by treatment.
A common strategy would choose promising outcome variables solely based on the smallest observed $p$-values, assuming the absence of unmeasured confounding; we refer to this method as the \emph{naive selector}.
In contrast, the generalized sensitivity value–based methods explicitly account for potential unmeasured confounding by selecting the subset $\cR$ that maximizes the value $\Gamma^*(\cR, 1)$ among all ${4 \choose 2} = 6$ candidate subsets of size 2. In words, this approach finds the subset of size 2 for which the finding that at least one outcome is affected by the treatment is the most robust to hidden bias. We refer to this selector as the \emph{$\Gamma^*(\cR, 1)$ selector}.

We set \(B=500\). Let
\(\rho_{12}\in\{-0.2,0,0.2\}\) denote the correlation between outcomes 1
and 2, while the correlation between outcomes 3 and 4 is fixed at \(0.2\).
The treatment-effect magnitude for outcomes 1 and 2 is calibrated
so that all four outcomes have similar average observed treated-minus-control
differences, making it difficult to separate true effects from spurious ones. Additional details on this data-generating process can be found in Section ~\ref{app:simulation} of the Supplementary Material.

To assess performance, we define a selector as \emph{successful} if it selects at least one outcome that is actually affected by the treatment. 
For each value of \(\rho_{12}\), we generate \(1{,}000\) datasets and report
the proportion of successful selections in
Table~\ref{tab:subsetselectors}. In each setting, the generalized
sensitivity-value selector has a higher success proportion than the naive selector, illustrating that using \(\Gamma^*(\cR,r)\) to guide subset selection more reliably identifies outcomes with genuine treatment effects, whereas the naive approach is more likely to select spurious signals.

\begin{table}[ht]
\centering
\begin{tabular}{|l|c|c|}
\hline
$\rho_{12}$ & Naive selector & $\Gamma^*(\mathcal{R},1)$ selector \\ 
\hline
-0.2 & 0.829 & 0.905 \\
\hline
0 & 0.810 & 0.893 \\
\hline
0.2 & 0.781 & 0.851 \\
\hline
\end{tabular}
\caption{Proportion of successful selections by two selectors.}
\label{tab:subsetselectors}
\end{table}
\vspace{-0.8cm}
\section{Application: Effects of CPEA exposure}\label{sec:application}

The Wisconsin Longitudinal Study (WLS) is a long-term investigation of individuals who graduated from Wisconsin high schools in 1957, with data collected in multiple follow-up waves.
For this analysis, we focus on respondents who participated in the 2004–2005 round. After excluding individuals with missing data on key covariates and outcomes, our working sample comprises $N=3469$ individuals. We investigate a total of 21 outcomes spanning multiple domains outlined in the motivating example in Section~\ref{sec:example}.

We use Rosenbaum's sensitivity analysis in
Example~\ref{ex:rosenbaum}, with the computations described in
Section~\ref{sec:computation} to analyze this data. 
We employ full matching \citep{han04} to reduce bias from 13 observable confounders. See Section ~\ref{app:application} of the Supplementary Material for the construction of the CPEA exposure indicator, an overview of the outcome variables, and the quality of the produced matched sets. We test Fisher's sharp null of no effect for each outcome variable. We use the Mantel–Haenszel statistic for binary outcomes, and for continuous endpoints we use an M-statistic with outer trimming constant 2.5 \citep{hub81, ros07}.


As a demonstration of the explorations enabled by our procedure, we first consider a \emph{post-hoc} selection strategy to identify a promising subset of the $21$ outcome variables for which the finding that half of them are actually affected by the treatment is robust to hidden bias. Once this subset is identified, we can leverage the simultaneity property of our procedure to provide a valid sensitivity set for the FDP for it. To proceed, we first compute the conventional $p$‐values for each of the 21 original outcome variables assuming no hidden bias ($\Gamma = 1$), and select those with $p \leq 0.05$. Denote this preliminary set by $\cR^0$, which contains 13 outcomes in our data. We next apply closed sensitivity analysis to the outcomes within $\cR^0$, computing the generalized sensitivity values $\Gamma^*(\cR, \lfloor |\mathcal{R}| / 2 \rfloor)$ for every subset $\cR \subseteq \cR^0$. 
We use this restricted search only to propose a candidate subset. All reported inferential statements below are then recomputed using the full collection of \(K=21\) outcomes, so that the simultaneous validity guarantee applies to the final reported subset. 
In particular, we found that the subset $\cR^*_1 = $ \{alcohol, closeness to spouse, anger and agreeableness\} exhibits a relatively large sensitivity value. To attain a valid sensitivity value for $\cR^*_1$, we apply our procedure using the original $21$ variables to this chosen subset. This yields a sensitivity value of $\Gamma^*(\cR^*_1, \lfloor |\mathcal{R}^*_1| / 2 \rfloor) = 1.46$. 
This means that, for every \(\Gamma<1.46\), the sensitivity set continues to support the claim that at least two of the four outcomes in \(\cR_1^*\) are affected by childhood abuse, implying that the conclusion remains robust until hidden bias is allowed to make two individuals within the same matched set differ in their odds of treatment assignment by a factor of \(1.46\).

To demonstrate the improved power of our method over the naive approach described in Section~\ref{subsec:naive}, we compare the generalized sensitivity values calculated using both methods for all subsets of size four. Across all subsets examined, our approach yielded larger generalized sensitivity values, aligning with our theoretical discussions in Section~\ref{subsec:naive}.
For instance, subset $\cR^*_1$ described above has a naive generalized sensitivity value of 1.42, which is 0.04 smaller than its exact value. For the subset $\cR^*_2 = $ \{neuroticism, closeness to spouse, anger and agreeableness\}, there is an even more notable gain: increasing from a naive generalized sensitivity value of 1.32 to 1.41 under our exact approach.
This improvement implies that our method can sustain the conclusion that at least half the outcomes in $\cR^*_2$ are true discoveries (i.e. truly affected by the treatment), even if hidden bias inflates the odds of differential treatment assignment within matched sets up to a factor of 1.41, compared to only 1.32 for the naive approach.

\section{Discussion}\label{sec:discussion}
We developed simultaneously valid sensitivity sets for the false discovery
proportion in observational studies with multiple outcomes. Our method applies beyond any particular study design or model for unmeasured confounding. Our proposed
generalized sensitivity value quantifies the robustness of the claim that at
least a specified number of outcomes in a subset are affected. We further
showed that the closed sensitivity analysis produces sharper sensitivity sets than the naive approach based on individual worst-case
\(p\)-values and improves statistical power. Focusing on Rosenbaum's sensitivity model for matched
observational studies, we formulated the calculation of the
FDP sensitivity sets as a sequence of mixed-integer quadratically constrained
programs and developed screening rules to reduce the required computation. Through simulation studies and an empirical application investigating the effects of childhood abuse, we demonstrated that our method provides greater statistical power, improved robustness to hidden biases, and significant computational advantages relative to naive or enumerative alternatives.

An immediate direction for future research is to develop complete
procedures under sensitivity models other than Rosenbaum's model. The
simultaneous validity result applies whenever a sensitivity model supplies a
set of candidate null distributions and valid local tests under each fixed
hidden bias configuration. Nevertheless, constructing and optimizing these
local tests is specific to the sensitivity model. For example, under a
marginal sensitivity model, the optimization must account for a common latent
propensity score function across outcomes, together with estimation of the
nominal propensity score and other nuisance quantities. Under a
factor-structured partial-\(R^2\) sensitivity model, computation would instead
involve optimization over the allowable association between treatment and
the latent factors, while accounting for estimation of the factor loadings
and residual covariance structure. Developing scalable algorithms and
establishing their inferential validity would turn the general applicability
of our framework into operational methods for a wider range of observational
studies.

Even within Rosenbaum's sensitivity model, an important extension would be to allow the magnitude of hidden bias to vary across matched sets.  Our current
implementation places the same upper bound on hidden bias in every matched
set, as in Example~\ref{ex:rosenbaum}. In practice, however, the degree of
residual confounding may be heterogeneous. \citet{wu25}, for example, propose
bounding a specified quantile of the matched-set-specific biases rather than
their maximum. This permits a limited fraction of matched sets to exhibit
larger hidden bias without requiring the same bound to hold uniformly.
Incorporating such constraints into the closed sensitivity analysis would
provide a more flexible assessment of robustness. The general construction of
FDP sensitivity sets would remain unchanged, but the computation would become
more difficult because quantile restrictions couple the hidden bias
configurations across matched sets.


\begingroup  %
\setlength{\bibsep}{0pt}
\bibliographystyle{apalike}
\bibliography{bibliography}
\endgroup

\newpage

\begin{center}
\textbf{\Huge{Supplementary Material}}
\end{center}

\appendix
\section{Additional derivations}\label{app:details}



\subsection{Derivation of the worst-case two-sided $p$-value $p_{k,\Gamma}^*$}\label{app:worst_p}

Given the probability assignment \(\brho\) and using the normal approximation of \(T_k\) in Equation~\eqref{eq:t-normal}, we obtain the normally approximated two-sided $p$-value under $\brho$ as:
\[
p_{k,\brho} = 2\left(1-\Phi\left(\left|\frac{T_k - \mu_{k,\brho}}{\sigma_{k,\brho}}\right|\right)\right) = 1 - F\left(\left(\frac{T_k - \mu_{k,\brho}}{\sigma_{k,\brho}}\right)^2\right),
\]
where \( F \) is the cumulative distribution function of the \(\chi_{1}^2\) distribution. 

For a given $\Gamma$, let \(\brho_{\Gamma}^*\) be the $\brho \in \mathcal{P}_\Gamma$ that maximizes \(p_{k,\brho}\), i.e.,
\[
\brho_{\Gamma}^* = \argmin_\adverrho \abs{\frac{T_k - \mu_{k,\brho}}{\sigma_{k,\brho}}}.
\]
Let $\mu_{k,\Gamma}^* = \mu_{k,\brho_{\Gamma}^*}$ and $\sigma_{k,\Gamma}^* = \sigma_{k,\brho_{\Gamma}^*}$ denote the expectation and standard deviation under this $\brho_{\Gamma}^*$, respectively.    
Then the $p$-value under this \(\brho_{\Gamma}^*\), which corresponds to the worst-case $p$-value under $\Gamma$, is given by:
\begin{equation}\label{eq:worst_p_nscreen}
    p_{k,\Gamma}^* = \max_\adverrho \;p_{k,\brho} = p_{k,\brho_{\Gamma}^*} = 2\left(1-\Phi\left(\left|\frac{T_k - \mu_{k,\Gamma}^*}{\sigma_{k,\Gamma}^*}\right|\right)\right).
\end{equation}
                 
Notice that, for a given $\Gamma$, when $T_k = \mu_{k,\Gamma}^*$ is attainable, $p_{k,\Gamma}^* = 1$. 
Moreover, since for $\Gamma' \geq \Gamma$, we have $\mathcal{P}_\Gamma \subseteq \mathcal{P}_{\Gamma'}$, therefore, $T_k = \mu_{k,\Gamma'}^*$ is also attainable, implying $p_{k,\Gamma'}^* = 1$.

\section{Proofs of results in Section~\ref{sec:SA_FDP}}
\subsection{Proof of Theorem~\ref{thm:sensitivity_sets_valid}}
\begin{proof}
Let
\[
\mathcal E_0
=
\left\{
V(\cR)\leq v_{\brho_0}(\cR)
\ \text{for every }\cR\subseteq[K]
\right\}.
\]
Applying the simultaneous FDP confidence result of \citet{goemansolari} to
the closed testing procedure under the true configuration \(\brho_0\) gives
\[
\P_{\brho_0}(\mathcal E_0)\geq 1-\alpha.
\]

Now let \(\Gamma\geq\Gamma_0\). By the nesting of the sensitivity models,
\(\cP_{\Gamma_0}\subseteq\cP_\Gamma\), and hence
\(\brho_0\in\cP_\Gamma\). Therefore, for every \(\cR\subseteq[K]\),
\[
v_{\brho_0}(\cR)
\leq
\max_{\brho\in\cP_\Gamma}v_{\brho}(\cR)
=
v_\Gamma^*(\cR).
\]
Consequently, on the single event \(\mathcal E_0\),
\[
V(\cR)\leq v_\Gamma^*(\cR)
\]
holds for every \(\cR\subseteq[K]\) and every
\(\Gamma\geq\Gamma_0\). Thus,
\[
\P_{\brho_0}\left\{
V(\cR)\leq v_\Gamma^*(\cR)
\ \text{for every }\cR\subseteq[K]
\text{ and every }\Gamma\geq\Gamma_0
\right\}
\geq 1-\alpha.
\]
For every nonempty \(\cR\), dividing by \(|\cR|\) gives the claimed
simultaneous sensitivity set for the FDP. The empty set has FDP zero and is
trivial.
\end{proof}

\subsection{Proof of Proposition~\ref{prop:collective_robustness}}
\begin{proof}
For any fixed \(\Gamma\), suppose that
\[
    v_\Gamma^*(\cR)>|\cR|-1.
\]
Since \(v_\Gamma^*(\cR)\le |\cR|\), this implies \(v_\Gamma^*(\cR)=|\cR|\). Hence
there exists some \(\brho\in\mathcal{P}_\Gamma\) such that
\(v_{\brho}(\cR)=|\cR|\), which can occur only if the full intersection hypothesis
\(H_\cR\) is not rejected by the closed testing procedure under \(\brho\). By
closed testing, if \(H_\cR\) is not rejected, then no elementary hypothesis
\(H_k\), \(k\in\cR\), is rejected under the same \(\brho\). Therefore,
\[
    v_\Gamma^*(\{k\})>0
    \quad\text{for every }k\in\cR.
\]
Thus
\[
\left\{\Gamma\ge1:v_\Gamma^*(\cR)>|\cR|-1\right\}
\subseteq
\bigcap_{k\in\cR}
\left\{\Gamma\ge1:v_\Gamma^*(\{k\})>0\right\}.
\]
Taking infima gives
\[
    \Gamma^*(\cR,|\cR|-1)
    \ge
    \max_{k\in\cR}\Gamma^*(\{k\},0).
\]
\end{proof}

\subsection{Proof of Proposition~\ref{prop:closed_sensitivity_representation}}

\begin{proof}
By definition of closed testing, $\widetilde{\phi}_{\cI,\brho}
=
\min_{\cJ\supseteq\cI}\phi_{\cJ,\brho}.$
We have
\[
\begin{aligned}
\min_{\brho\in\cP_\Gamma}
\widetilde{\phi}_{\cI,\brho}
&=
\min_{\brho\in\cP_\Gamma}
\min_{\cJ\supseteq\cI}\phi_{\cJ,\brho} \\
&=
\min_{\cJ\supseteq\cI}
\min_{\brho\in\cP_\Gamma}\phi_{\cJ,\brho} \\
&=
\min_{\cJ\supseteq\cI}\phi_{\cJ,\Gamma}^*.
\end{aligned}
\]
It follows that
\[
\begin{aligned}
v_\Gamma^*(\cR)
&=
\max_{\brho\in\cP_\Gamma}
\max_{\cI\subseteq\cR}
|\cI|
\left(1-\widetilde{\phi}_{\cI,\brho}\right) \\
&=
\max_{\cI\subseteq\cR}
|\cI|
\left(
1-
\min_{\brho\in\cP_\Gamma}
\widetilde{\phi}_{\cI,\brho}
\right) \\
&=
\max_{\cI\subseteq\cR}
|\cI|
\left(
1-\min_{\cJ\supseteq\cI}
\phi_{\cJ,\Gamma}^*
\right).
\end{aligned}
\]
Using
\[
1-\min_{\cJ\supseteq\cI}\phi_{\cJ,\Gamma}^*
=
\max_{\cJ\supseteq\cI}
\left(1-\phi_{\cJ,\Gamma}^*\right),
\]
we obtain
\[
\begin{aligned}
v_\Gamma^*(\cR)
&=
\max_{\cI\subseteq\cR}
\max_{\cJ\supseteq\cI}
|\cI|
\left(1-\phi_{\cJ,\Gamma}^*\right) \\
&=
\max_{\cJ\subseteq[K]}
\max_{\cI\subseteq\cR\cap\cJ}
|\cI|
\left(1-\phi_{\cJ,\Gamma}^*\right) \\
&=
\max_{\cJ\subseteq[K]}
|\cJ\cap\cR|
\left(1-\phi_{\cJ,\Gamma}^*\right).
\end{aligned}
\]
The last equality follows because
\(1-\phi_{\cJ,\Gamma}^*\geq0\), so for each fixed \(\cJ\), the inner
maximum is attained at \(\cI=\cR\cap\cJ\).

Finally, the minimax inequality in \eqref{eq:minimax} gives
\[
\phi_{\cJ,\Gamma}^{\mathrm{naive}}
\leq
\phi_{\cJ,\Gamma}^*
\]
for every nonempty \(\cJ\subseteq[K]\). Therefore,
\[
|\cJ\cap\cR|
\left(1-\phi_{\cJ,\Gamma}^*\right)
\leq
|\cJ\cap\cR|
\left(1-\phi_{\cJ,\Gamma}^{\mathrm{naive}}\right)
\]
for every \(\cJ\). Taking the maximum over \(\cJ\) proves the desired
inequality.
\end{proof}
\subsection{Proof of Proposition~\ref{prop:fdp_gain_nonconsonance}}
\begin{proof}
The argument follows \citet{goemansolari}.
Fix \(\cR\subseteq[K]\) and write \(U=U_\Gamma(\cR)\). By
Proposition~\ref{prop:closed_sensitivity_representation},
\(v_\Gamma^*(\cR)\) is the largest cardinality of a set
\(\cI\subseteq\cR\) whose intersection hypothesis is not rejected by
the closed sensitivity analysis.

Suppose that \(H_{\cI}\) is not rejected. Then there exists some
\(\cJ\supseteq\cI\) for which \(\phi_{\cJ,\Gamma}^*=0\). For every
\(k\in\cI\), the same set \(\cJ\) contains \(\{k\}\), and therefore
prevents the closed sensitivity analysis from rejecting \(H_{\{k\}}\).
Hence \(k\in U\), so \(\cI\subseteq U\). It follows that
\[
v_\Gamma^*(\cR)\leq |U|.
\]

If \(H_U\) is not rejected, then \(U\) itself is among the sets over
which the maximum defining \(v_\Gamma^*(\cR)\) is taken. Consequently,
\[
v_\Gamma^*(\cR)\geq |U|,
\]
and hence \(v_\Gamma^*(\cR)=|U|\). Conversely, if \(H_U\) is rejected,
then every nonrejected intersection \(\cI\subseteq\cR\) is a strict
subset of \(U\). Therefore,
\[
v_\Gamma^*(\cR)<|U|.
\]
This proves the stated equivalence.

Finally, every elementary hypothesis \(H_{\{k\}}\), \(k\in U\), is
unrejected by the definition of \(U\). Thus, when \(H_U\) is rejected,
the intersection hypothesis is rejected without the rejection of any
of its elementary hypotheses. The rejection is therefore
non-consonant.

\end{proof}
\section{Proofs of results in Section~\ref{sec:computation}}\label{app:comp_proofs}
\subsection{Proof of Theorem~\ref{thm:normal_validity}}
\begin{proof}
Define
\[
\Delta_B=
\max_{k\in[K]}\sup_{\brho\in\cP_{\Gamma_0}}
\frac{
\sum_{b=1}^B
\E_{\brho}\!\left[
\left|
\sum_{i\in\mathcal M_b}(Z_i-\varrho_i)q_{ik}
\right|^3
\,\middle|\,\cF,\cZ
\right]
}{
\sigma_{k,\brho}^3
}.
\]
For \(k\in[K]\) and \(\brho\in\cP_{\Gamma_0}\), write
\[
Z_{k,\brho}
=
\frac{T_k-\mu_{k,\brho}}{\sigma_{k,\brho}}.
\]
By the Berry--Esseen inequality, there is a universal constant
\(C>0\) such that
\[
\sup_{x\in\mathbb R}
\left|
\P_{\brho}(Z_{k,\brho}\leq x)-\Phi(x)
\right|
\leq C\Delta_B
\]
uniformly over \(k\in[K]\) and \(\brho\in\cP_{\Gamma_0}\).

Let
\[
p_{k,\brho}^{\mathrm N}
=
2\left\{1-\Phi\bigl(|Z_{k,\brho}|\bigr)\right\}
\]
denote the normally approximated two-sided \(p\)-value. Whenever
\(H_k\) is true, for every \(u\in(0,1)\),
\[
\P_{\brho}\left(p_{k,\brho}^{\mathrm N}\leq u\right)
\leq u+2C\Delta_B.
\]

Consequently, for any nonempty \(\cI\subseteq[K]\) such that
\(H_{\cI}\) is true, the Bonferroni local test satisfies
\[
\begin{aligned}
\P_{\brho_0}\left(\phi_{\cI,\brho_0}=1\right)
&=
\P_{\brho_0}\left\{
\bigcup_{k\in\cI}
\left(
p_{k,\brho_0}^{\mathrm N}
\leq\frac{\alpha}{|\cI|}
\right)
\right\}\\
&\leq
\sum_{k\in\cI}
\P_{\brho_0}\left(
p_{k,\brho_0}^{\mathrm N}
\leq\frac{\alpha}{|\cI|}
\right)\\
&\leq
\sum_{k\in\cI}
\left\{
\frac{\alpha}{|\cI|}+2C\Delta_B
\right\}\\
&=
\alpha+2C|\cI|\Delta_B
\leq
\alpha+2CK\Delta_B.
\end{aligned}
\]
Let \(\cI_0\) denote the set of true null hypotheses. Taking
\(\cI=\cI_0\) gives
\[
\P_{\brho_0}\left(\phi_{\cI_0,\brho_0}=1\right)
\leq
\alpha+2CK\Delta_B.
\]

Because \(\brho_0\in\cP_{\Gamma_0}\),
\[
\phi_{\cI_0,\Gamma_0}^*
=
\min_{\brho\in\cP_{\Gamma_0}}\phi_{\cI_0,\brho}
\leq
\phi_{\cI_0,\brho_0}.
\]
Hence,
\[
\P_{\brho_0}\left(\phi_{\cI_0,\Gamma_0}^*=1\right)
\leq
\alpha+2CK\Delta_B.
\]
If \(\phi_{\cI_0,\Gamma_0}^*=0\), then \(H_{\cI_0}\) is not rejected
by closed testing, and consequently
\[
v_{\Gamma_0}^*(\cR)
\geq|\cR\cap\cI_0|
=
V(\cR)
\]
simultaneously for every \(\cR\subseteq[K]\).

Moreover, for every \(\Gamma\geq\Gamma_0\),
\(\cP_{\Gamma_0}\subseteq\cP_\Gamma\), so
\[
v_{\Gamma}^*(\cR)\geq v_{\Gamma_0}^*(\cR).
\]
Therefore,
\[
\P_{\brho_0}\left\{
V(\cR)\leq v_\Gamma^*(\cR)
\ \text{for every }\cR\subseteq[K]
\text{ and every }\Gamma\geq\Gamma_0
\right\}
\geq
1-\alpha-2CK\Delta_B.
\]
Since \(K\) is fixed and \(\Delta_B\to0\), the right-hand side is
\(1-\alpha-o(1)\). Dividing by \(|\cR|\) for every nonempty
\(\cR\) gives the stated sensitivity sets for the FDP.
\end{proof}
\subsection{Proof of Theorem~\ref{thm:IP}}\label{app:proof_thmIP}
\begin{proof}
Write
\[
t_{bk}=\sum_{i\in\mathcal M_b}Z_iq_{ik},
\qquad
T_k=\sum_{b=1}^B t_{bk}, \qquad Q_{bk}=\max_{i\in\mathcal M_b}|q_{ik}|.
\]
Since each matched set contains one treated unit,
\(-Q_{bk}\le t_{bk}\le Q_{bk}\). The 
condition $\limsup_{B\to\infty}
\max_{k\in[K]}
\frac{1}{B}\sum_{b=1}^B \max_{i\in\mathcal M_b}|q_{ik}|^2
<\infty$ implies that, for some \(C<\infty\) and all sufficiently
large \(B\),
\[
\max_{k\in[K]}\sum_{b=1}^B Q_{bk}^2\le CB.
\]

For every \(\brho\in\cP_\Gamma\),
\[
\begin{aligned}
\sigma_{k,\brho}^2
&=
\sum_{b=1}^B
\Var_{\brho}(t_{bk}\mid\cF,\cZ)\\
&\le
\sum_{b=1}^B
\E_{\brho}(t_{bk}^2\mid\cF,\cZ)
\le
\sum_{b=1}^B Q_{bk}^2
\le CB.
\end{aligned}
\]
This bound is uniform over \(\brho\), and therefore:
\[
\max_{k\in[K]}
\frac{(\sigma_{k,\Gamma}^*)^2}{B}
\le C.
\]

Define
\[
\mathcal E_B
=
\bigcap_{k=1}^K
\left[
\left\{\mu_{k,\Gamma}^*=T_k\right\}
\cup
\left\{
\frac{|\mu_k-\mu_{k,\Gamma}^*|}{B}\ge\epsilon
\right\}
\right]
\]
and
\[
\mathcal H_B
=
\left\{
\max_{k\in[K]}|T_k-\mu_k|<\frac{\epsilon B}{2}
\right\}.
\]
Because \(K\) is fixed, the assumed condition and a union bound imply
\(\P(\mathcal E_B)\to1\), while the assumed concentration condition implies
\(\P(\mathcal H_B)\to1\).

On \(\mathcal E_B\cap\mathcal H_B\), consider any \(k\in[K]\). If
\(\mu_{k,\Gamma}^*=T_k\), then \(p_{k,\Gamma}^*=1\). Otherwise,
\[
|T_k-\mu_{k,\Gamma}^*|
\ge
|\mu_k-\mu_{k,\Gamma}^*|-|T_k-\mu_k|
\ge
\frac{\epsilon B}{2}.
\]
It follows that
\[
\left|
\frac{T_k-\mu_{k,\Gamma}^*}
     {\sigma_{k,\Gamma}^*}
\right|
\ge
\frac{\epsilon\sqrt B}{2\sqrt C},
\]
and therefore
\[
p_{k,\Gamma}^*
=
2\left[
1-\Phi\left(
\left|
\frac{T_k-\mu_{k,\Gamma}^*}
     {\sigma_{k,\Gamma}^*}
\right|
\right)
\right]
\le
2\left\{
1-\Phi\left(\frac{\epsilon\sqrt B}{2\sqrt C}\right)
\right\}
\le
\exp\left(-\frac{\epsilon^2B}{8C}\right).
\]
Since \(K\) is fixed, the last expression is no larger than
\(\alpha/K\) for all sufficiently large \(B\). Thus, on
\(\mathcal E_B\cap\mathcal H_B\), every \(p_{k,\Gamma}^*\) is either
equal to \(1\) or no larger than \(\alpha/K\). Consequently,
\[
\begin{aligned}
\P\left(
\bigcup_{k=1}^K
\left\{
\frac{\alpha}{K}<p_{k,\Gamma}^*\le\alpha
\right\}
\right)
\le
\P(\mathcal E_B^c)+\P(\mathcal H_B^c)\longrightarrow0.
\end{aligned}
\]
\end{proof}

\subsection{Conditions and proof of Remark~\ref{rem:IP_growingK}}
\label{app:proof_growingK}

Let \(K=K_B\). The extension in Remark~\ref{rem:IP_growingK}
holds under the following conditions. First,
\[
\max_{k\in[K_B]}
\frac{|T_k-\mu_k|}{B}
\overset{p}{\longrightarrow}0.
\]
Second,
\[
\limsup_{B\to\infty}
\max_{k\in[K_B]}
\frac{1}{B}
\sum_{b=1}^B
\max_{i\in\mathcal M_b}|q_{ik}|^2
<\infty.
\]
Third, for some common constant \(\epsilon>0\),
\[
\P\left(
\bigcap_{k=1}^{K_B}
\left[
\left\{\mu_{k,\Gamma}^*=T_k\right\}
\cup
\left\{
\frac{|\mu_k-\mu_{k,\Gamma}^*|}{B}
\ge\epsilon
\right\}
\right]
\right)
\longrightarrow1.
\]
Finally, suppose that
\[
\log K_B=o(B).
\]
Under these conditions,
\[
\P\left(
\bigcup_{k=1}^{K_B}
\left\{
\frac{\alpha}{K_B}<p_{k,\Gamma}^*\le\alpha
\right\}
\right)
\longrightarrow0.
\]

\begin{proof}
Define
\[
\mathcal A_B
=
\left\{
\max_{k\in[K_B]}
|T_k-\mu_k|
<
\frac{\epsilon B}{2}
\right\}
\]
and
\[
\mathcal E_B
=
\bigcap_{k=1}^{K_B}
\left[
\left\{\mu_{k,\Gamma}^*=T_k\right\}
\cup
\left\{
|\mu_k-\mu_{k,\Gamma}^*|
\ge\epsilon B
\right\}
\right].
\]
By the first and third conditions,
\[
\P(\mathcal A_B)\longrightarrow1
\qquad\text{and}\qquad
\P(\mathcal E_B)\longrightarrow1.
\]
Consequently,
\[
\P(\mathcal A_B\cap\mathcal E_B)\longrightarrow1.
\]

Let
\[
Q_{bk}=\max_{i\in\mathcal M_b}|q_{ik}|.
\]
The uniform score bound implies that, for some \(C<\infty\) and all
sufficiently large \(B\),
\[
\max_{k\in[K_B]}
\sum_{b=1}^B Q_{bk}^2
\le CB.
\]
Because the contribution from matched set \(b\) is bounded in absolute
value by \(Q_{bk}\), its variance under any admissible configuration is
at most \(Q_{bk}^2\). Therefore,
\[
\max_{k\in[K_B]}(\sigma_{k,\Gamma}^*)^2
\le
\max_{k\in[K_B]}
\sum_{b=1}^B Q_{bk}^2
\le CB.
\]

Now consider the event \(\mathcal A_B\cap\mathcal E_B\). For any
\(k\in[K_B]\), if \(\mu_{k,\Gamma}^*=T_k\), then $p_{k,\Gamma}^*=1.$
Otherwise,
\[
\begin{split}
|T_k-\mu_{k,\Gamma}^*|
&\ge
|\mu_k-\mu_{k,\Gamma}^*|-|T_k-\mu_k| \\
&\ge
\frac{\epsilon B}{2}.
\end{split}
\]
It follows that
\[
\begin{split}
p_{k,\Gamma}^*
&\le
2\left\{
1-\Phi\left(
\frac{\epsilon\sqrt B}{2\sqrt C}
\right)
\right\} \\
&\le
\exp\left(-\frac{\epsilon^2B}{8C}\right).
\end{split}
\]
Since \(\log K_B=o(B)\),
\[
\exp\left(-\frac{\epsilon^2B}{8C}\right)
\le
\frac{\alpha}{K_B}
\]
for all sufficiently large \(B\). Hence, on
\(\mathcal A_B\cap\mathcal E_B\), every \(p_{k,\Gamma}^*\) is either
equal to \(1\) or no larger than \(\alpha/K_B\). Therefore,
\[
\begin{split}
&\P\left(
\bigcup_{k=1}^{K_B}
\left\{
\frac{\alpha}{K_B}<p_{k,\Gamma}^*\le\alpha
\right\}
\right) \\
&\qquad\le
\P(\mathcal A_B^c)+\P(\mathcal E_B^c)
\longrightarrow0.
\end{split}
\]
The conclusion concerning the integer program follows from the
singleton screening rule.
\end{proof}

\section{Additional runtime improvements}\label{app:comp}

This section presents additional computational optimizations for computing \(v_{\Gamma}^*(\cR)\) in Algorithm~\ref{alg:IP}. These improvements primarily leverage the worst-case \(p\)-values \(p_{1,\Gamma}^*, \ldots, p_{K,\Gamma}^*\).

\subsection{Refining the search range for \texorpdfstring{\(r\)}{r}}
Algorithm~\ref{alg:IP} searches for the possible value of \(v_{\Gamma}^*(\cR)\) from \(|\cR|\) down to 1. We propose upper and lower bounds on \(v_{\Gamma}^*(\cR)\) that allow us to restrict this search to the narrower range \(\{v_{\Gamma}^*(\cR)_{\text{up}}, \,v_{\Gamma}^*(\cR)_{\text{up}} -1, \,\ldots, \, v_{\Gamma}^*(\cR)_{\text{lo}}\}\). 

\paragraph{Upper bound \(\displaystyle v_{\Gamma}^*(\cR)_{\text{up}}\).}
Recall the naive approach from Section~\ref{subsec:naive}, which applies the Holm-Bonferroni procedure to the worst-case \(p\)-values \(p_{k,\Gamma}^*\). Let
\[
  \cR_\Gamma^{\text{naive}}
  \;=\;
  \bigl\{\,k : \tilde{\phi}_{k,\Gamma}^{\text{naive}} = 1\,\bigr\}
\]
be the set of elementary hypotheses rejected by the naive procedure. From Section~\ref{subsec:naive}, the quantity \(v_\Gamma^{\text{naive}}(\cR)\) computed via the naive approach upper-bounds the exact value: $  v_{\Gamma}^*(\cR) \leq v_\Gamma^{\text{naive}}(\cR)$, which leads us to set $  v_{\Gamma}^*(\cR)_{\text{up}} = v_\Gamma^{\text{naive}}(\cR).$

\paragraph{Lower bound \(\displaystyle v_{\Gamma}^*(\cR)_{\text{lo}}\).}
If there exists some \(k \in \cR\) with \(p_{k,\Gamma}^* > \alpha\), then \(\phi_{k,\Gamma}^* = 0\) and consequently \(\tilde{\phi}_{k,\Gamma}^* = 0\), which follows immediately that $v_{\Gamma}^*(\cR) \geq 1$.
Thus, for sets \(\cR\) containing such a hypothesis \(k\), we have a non-trivial lower bound of \(\displaystyle 1\). If no hypothesis in \(\cR\) satisfies \(p_{k,\Gamma}^* > \alpha\), the search lower bound remains 0.

\subsection{Reducing candidate supersets in the integer program}

Even with refined bounds for \(r\), enumerating all supersets \(\cJ\) of each $\cI$ can be computationally prohibitive. We therefore introduce two further optimizations that reduce the candidate \(\cJ\) sets needed to be evaluated:

\paragraph{Eliminating subsets containing naive rejections.}
Under the naive procedure, if an elementary hypothesis \(k\) is rejected (\(\tilde{\phi}_{k,\Gamma}^{\text{naive}} = 1\)), then for any set \(\cJ\) containing \(k\), we have \(\phi_{\cJ,\Gamma}^{\text{naive}} = 1\). By Equation~\eqref{eq:minimax}, it follows that \(\phi_{\cJ,\Gamma}^{\text{naive}} \leq \phi_{\cJ,\Gamma}^{*}\) for all \(\cJ\), implying \(\phi_{\cJ,\Gamma}^{*} = 1\). Consequently, no such \(\cJ\subseteq[K]\) can remain un-rejected in the closed sensitivity analysis. Hence, when evaluating subsets \(\cJ\) of fixed cardinality \(v\) in the integer program, we exclude any hypothesis already rejected by the naive procedure by setting \(\theta_k=0\) for all \(k \in \cR_\Gamma^{\text{naive}}\).

\paragraph{Excluding hypotheses with small worst-case \texorpdfstring{\(p\)-values}{p-values}.}
When evaluating all candidate sets \(\cJ\) of size \(v\) in the integer program, if there exists a \(k^\star \in \cJ\) with \(p_{k^\star,\Gamma}^* \le \alpha/v\), then
\begin{equation*}
\phi_{\cJ,\Gamma}^{*} = 
\mathbf{1}\left\{ \max_{ \brho \in \mathcal{P}_\Gamma } \min_{k \in \cJ} \,p_{k,\brho} \leq \frac{\alpha}{|\cJ|} \right\} \geq \mathbf{1}\left\{ \max_{ \brho \in \mathcal{P}_\Gamma }  \,p_{k^\star,\brho} \leq \frac{\alpha}{|\cJ|} \right\} = \mathbf{1}\left\{  p_{k^\star,\Gamma}^* \leq \frac{\alpha}{|\cJ|} \right\} = 1. 
\end{equation*}
That is, the local test for \(H_{\cJ}\) is guaranteed to reject in the closed sensitivity analysis, and there is no need to consider such \(\cJ\) further. Hence, we set \(\theta_{k} = 0\) for every \(k\) with \(p_{k,\Gamma}^* \le \alpha/v\), effectively excluding those hypotheses from all candidate sets.

By incorporating the above bounds and subset-elimination strategies, we arrive at the improved integer program in Algorithm~\ref{alg:IP_improved}.  Overall, these enhancements reduce the computational complexity, especially for large-scale problems. 
\begin{algorithm}[H]
\caption{Improved Integer Program for Computing \(v_{\Gamma}^*(\mathcal{R})\)}
\label{alg:IP_improved}
\begin{mdframed}
\textbf{For } \( r \in \{\,v_{\Gamma}^*(\cR)_{\text{up}}, \ldots, v_{\Gamma}^*(\cR)_{\text{lo}}\}\):
\begin{align*}
    \text{For } v \in \{r,\ldots,K\}: \quad & \\
    & \text{minimize } \quad y,\\
    & \text{subject to:}\\
    & \quad \boldsymbol{\varrho} \in \mathcal{P}_\Gamma,\\
    & \quad \sum_{k \in \mathcal{R}} \theta_k = r,\\
    & \quad \sum_{k \in [K]} \theta_k = v,\\
    & \quad \theta_k = 0 \quad\text{if } p_{k,\Gamma}^* \le \alpha/v,\\
    & \quad \theta_k = 0 \quad\text{if } k \in \cR_\Gamma^{\text{naive}},\\
    & \quad y \ge \zeta_{k}(\brho;\alpha/v) \quad \text{if } \theta_k = 1,\\
    & \quad y < 0,\\
    & \quad \theta_k \in \{0,1\}\;\text{for all }k,\\
    & \text{If feasible: } \text{return } r. \\
    & \text{Otherwise: continue.}
\end{align*}
\textbf{Return } 0.
\end{mdframed}
\end{algorithm}

\section{Additional details for the simulations in
Section~\ref{subsec:gsv_subset_selection}}
\label{app:simulation}

This appendix describes the data-generating process for the 
simulation study in Section~\ref{subsec:gsv_subset_selection}. The potential outcomes
are generated according to Equation~\eqref{eq:simu_setting}, with
\[
    \btau
    =
    (\tau,\tau,0,0)^\top
    \qquad\text{and}\qquad
    \Sigma
    =
    \begin{pmatrix}
        1 & \rho_{12} & 0 & 0 \\
        \rho_{12} & 1 & 0 & 0 \\
        0 & 0 & 1 & 0.2 \\
        0 & 0 & 0.2 & 1
    \end{pmatrix},
\]
where \(\rho_{12}\in\{-0.2,0,0.2\}\). Thus, outcomes 1 and 2 have
constant treatment effect \(\tau=0.25\), whereas outcomes 3 and 4 satisfy their
sharp null hypotheses.

We generate treatment assignments using outcomes 3 and 4. For each unit
\(i\in[N]\), define $S_i=r_{Ci3}+r_{Ci4}.$
Because \(\tau_3=\tau_4=0\), we also have $S_i=r_{Ti3}+r_{Ti4}=R_{i3}+R_{i4}.$
For each matched pair \(\mathcal M_b=\{i,i'\}\), set
\(\Gamma_{\mathrm{bias}}=1.75\) and define
\[
    \varrho_i
    =
    \P(Z_i=1\mid\cF,\cZ)
    =
    \begin{cases}
        \displaystyle
        \frac{\Gamma_{\mathrm{bias}}}
             {1+\Gamma_{\mathrm{bias}}},
        & S_i>S_{i'},\\[8pt]
        \displaystyle\frac12,
        & S_i=S_{i'},\\[8pt]
        \displaystyle
        \frac{1}
             {1+\Gamma_{\mathrm{bias}}},
        & S_i<S_{i'},
    \end{cases}
    \qquad
    \varrho_{i'}=1-\varrho_i.
\]
We then generate
\[
    Z_i\sim\operatorname{Bernoulli}(\varrho_i),
    \qquad
    Z_{i'}=1-Z_i,
\]
independently across matched pairs.

Under this assignment mechanism, the unit with the larger sum of potential
outcomes 3 and 4 is more likely to receive treatment. Moreover,
\[
    \frac{\max_{i\in\mathcal M_b}\varrho_i}
         {\min_{i\in\mathcal M_b}\varrho_i}
    =
    \Gamma_{\mathrm{bias}},
\]
except in the event of a tie. Thus, the true assignment configuration
\(\brho_0=(\varrho_1,\ldots,\varrho_N)^\top\) belongs to
\(\cP_{\Gamma_{\mathrm{bias}}}\) under Rosenbaum's sensitivity model in
Example~\ref{ex:rosenbaum}.

Although outcomes 3 and 4 have no causal effects, their dependence on treatment
assignment produces positive observed treated-minus-control differences. With \(\tau=0.25\) and \(\Gamma_{\mathrm{bias}}=1.75\), the four outcomes have approximately equal average observed treated-minus-control
differences, making genuine and spurious effects difficult to distinguish.
\section{Additional details for the data analysis in Section~\ref{sec:application}}\label{app:application}
To define CPEA exposure, we use six questions answered by each WLS respondent—three regarding their father and three regarding their mother—on whether, before age 18, the parent (i) slapped, shoved, or threw things, (ii) insulted or swore, or (iii) treated them in a way they would now consider physical or emotional abuse. The response options were “a lot,” “some,” “a little,” or “not at all.” We label an individual as CPEA-exposed (\(Z = 1\)) if they reported experiencing any level of abuse (even “a little”) on at least one item for either parent. Conversely, individuals who answered “not at all” to all questions for both parents are labeled as CPEA-unexposed (\(Z = 0\)). Based on this definition, 1542 individuals are classified as exposed, while 1927 are classified as unexposed.

In line with the broad literature on CPEA (see Section~\ref{sec:example}), we consider 21 outcomes that prior studies have linked to CPEA. These include educational and economic attainment (college attendance, income), health behaviors (smoking, alcohol use), relationship quality (closeness with spouse, social support), physical health (obesity), emotional traits (anger, anxiety), and validated psychological scales covering both Ryff’s six-factor well-being \citep{ryff1989happiness} and the Big Five personality factors \citep{mccrae1992introduction}. Collectively, these measures capture the wide-ranging potential impacts of CPEA on outcomes in adulthood.

\begin{figure}[h!]
\begin{center}
\includegraphics{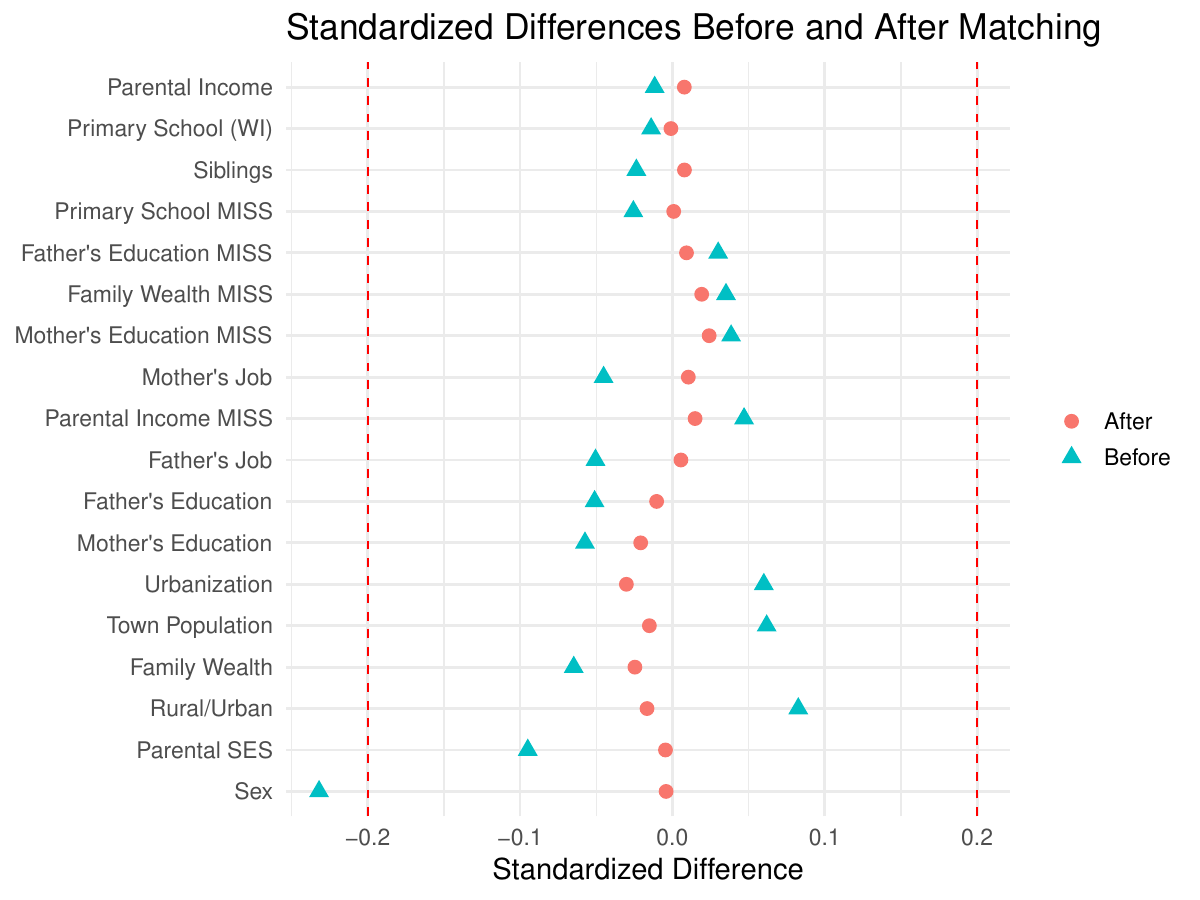}
\caption{\small
\textbf{Covariate Imbalances Before and After Matching.} The dotplot shows the standardized differences for each covariate before and after matching. The vertical dotted lines mark \(-0.2\) and \(0.2\), corresponding to the commonly used absolute-standardized-difference threshold of \(0.2\)  \citep{sil01}. After matching, the largest absolute standardized difference was 0.03, indicating that the matching procedure substantially reduced covariate imbalance.}
\label{fig:balance}
\end{center}
\end{figure}

We employ full matching \citep{ros91, han04} to control for observed covariates between these two groups. Each matched set can contain up to five total subjects, such that a set could include either 1 CPEA-exposed and at most 4 CPEA-unexposed individuals, or 1 CPEA-unexposed and at most 4 CPEA-exposed individuals. In total, we match on 13 pre-treatment covariates believed to predict CPEA; these covariates are listed in Figure~\ref{fig:balance}. Five of the covariates contain missing values, so we also incorporate 4 missingness indicators to balance both the observed covariates and the pattern of missingness \citep{ros84,designofobs}. Specifically, we use a rank-based Mahalanobis distance with a propensity score caliper of 0.08, where propensity scores are estimated via logistic regression \citep[see][Section~8.3]{designofobs}. 

Figure~\ref{fig:balance} shows the standardized differences before and after matching, highlighting a large initial imbalance between CPEA-exposed and CPEA-unexposed individuals for some key variables. After matching, these discrepancies are substantially reduced: the largest absolute standardized difference is 0.03, well below the commonly cited threshold of 0.2 \citep{sil01}. Details for calculating standardized differences before and after full matching can be found in \citet{stu08} and \citet[][Section 9.1]{designofobs}.

\end{document}